\documentclass[3p,times]{elsarticle}

\usepackage{ecrc}

\volume{197}

\firstpage{1}

\journalname{Chaos, Solitons \& Fractals}

\runauth{Hasmi \& Susanto}

\jid{procs}

\jnltitlelogo{Chaos, Solitons \& Fractals}

\CopyrightLine{2025}{Published by Elsevier Ltd.}

\usepackage{amssymb}
\usepackage{color}

\usepackage[figuresright]{rotating}
\usepackage{amsmath}
\usepackage{subfigure}

\usepackage{comment}

\begin{document}

\begin{frontmatter}



\dochead{}

\title{Supratransmission of rogue wave-like breathing solitons}

\author[au1]{Abrari Noor Hasmi}
\ead{100060615@ku.ac.ae}

\author[au1,au2]{Hadi Susanto\corref{cor1}}
\ead{hadi.susanto@yandex.com}

\address[au1]{Department of Mathematics, Khalifa University of Science and Technology, PO Box 127788, Abu Dhabi, United Arab Emirates}
  \cortext[cor1]{Corresponding author}
\address[au2]{Department of Mathematics, Faculty of Mathematics and Natural Sciences, Universitas Indonesia,\\ Gedung D Lt.\ 2 FMIPA Kampus UI Depok, 16424, Indonesia}

\begin{abstract}
Supratransmission is a fascinating and counterintuitive nonlinear wave phenomenon that enables energy transmission through frequency band gaps. Recent studies have suggested that supratransmission in a damped-driven Klein-Gordon equation can lead to the formation of rogue waves. In this work, we reduce the same model to a damped and driven discrete nonlinear Schr\"odinger equation and demonstrate that the claimed rogue waves are unstable solitons exhibiting breathing dynamics. Through extensive numerical simulations, we thoroughly investigate and analyze the behavior of solitons and supratransmission in these lattices. Additionally, we perform asymptotic analysis to explain the underlying mechanism responsible for supratransmission. Our findings provide new insights into the dynamics of nonlinear waves and their stability in such systems.
\end{abstract}


\end{frontmatter}



\section{Introduction}
\label{sec1}

Supratransmission is an intriguing nonlinear wave phenomenon that challenges the predictions of linear wave theory. According to linear theory, waves with frequencies within band gaps cannot transmit energy. However, in nonlinear systems, when the wave amplitude exceeds a critical threshold, the system's nonlinear properties enable energy to tunnel through the band gap, giving rise to supratransmission. This phenomenon was first discovered and reported by Leon and co-workers in a series of pioneering studies \cite{geniet2002energy,geniet2003nonlinear,leon2003nonlinear}.

The nonlinear breaching of forbidden energy bands is not only a fascinating theoretical concept but also has significant practical applications across various fields (see, e.g., the review \cite{zakharov2023effect}). Experimental evidence of supratransmission has been observed in electrical transmission lines \cite{bodo2010klein,koon2014experimental,tao2012experimental}. Similar phenomena have also been reported in harmonically driven monoatomic granular chains composed of a finite number of particles \cite{lydon2015nonlinear}. Additionally, controllable supratransmission has been demonstrated in mechanical metastructures with complex folding geometries, such as Miura origami \cite{zhang2020programmable} and Kresling origami \cite{wang2023highly} designs. Recent work by one of the current authors has further revealed intriguing wave transmission inhibition and supratransmission behaviors in flat and nearly flat frequency bands \cite{susanto2023surge}. Among its potential applications, supratransmission has been proposed for use in binary signal-based information transmission systems \cite{macias2007application}.

Recently, Motcheyo et al.\ \cite{motcheyo2022supratransmission,motcheyo2024nonlinear} reported an intriguing supratransmission phenomenon involving breathing structures, which they identified as rogue waves. Rogue waves are exceptionally large, spontaneous, and rare wave events that can appear without warning \cite{akhmediev2009waves}. Once considered maritime myths, such waves have now been observed across various physical systems \cite{onorato2013rogue,dudley2019rogue}. A widely studied prototype is the Peregrine soliton--a rational solution to the nonlinear Schr\"odinger equation that is localized in both space and time on an unstable continuous-wave background \cite{peregrine1983water}. This localization arises due to focusing modulational instability. Alternatively, rogue waves may emerge from a cascade of inelastic quasi-soliton collisions in generalized nonlinear Schr\"odinger equations, where resonant-like energy transfer produces large, localized structures \cite{eberhard2017rogue,peng2016rogue,gelash2018strongly}. Some studies argue that rogue waves are rare but statistically expected events within linear Gaussian wave theory, requiring no special nonlinear mechanisms \cite{teutsch2020statistical,dematteis2018rogue}.

The findings in \cite{motcheyo2024nonlinear} suggest that supratransmission could serve as a mechanism for rogue wave generation, offering new insight into their origin. However, no precise definition of rogue waves was provided. Given that the authors cite works involving Peregrine solitons, we reasonably interpret their use of the term ``rogue wave'' as referring to spatiotemporally localized Peregrine solitons. The primary objective of this study is to critically revisit the results in \cite{motcheyo2024nonlinear} and reinterpret the observed ``rogue waves'' within the framework of nonlinear lattice supratransmission. Our analysis reveals that the breathing structures they describe are rather unstable solitons whose transient dynamics resemble rogue wave behavior.

The structure of this paper is as follows. In Section~\ref{sec2}, we introduce the governing equations that form the basis of our model. Motcheyo et al.\ \cite{motcheyo2024nonlinear} used a damped-driven Klein-Gordon equation as their primary framework. Here, we reduce this equation to a driven and damped discrete nonlinear Schr\"odinger equation, which serves as the foundation for our analysis. Section~\ref{sec3} describes the numerical methods employed to simulate the supratransmission phenomenon. We present results illustrating supratransmission in two distinct regimes: small and large driving frequencies, corresponding to strongly and weakly coupled lattices, respectively. In Section~\ref{sec:thr_edge_analysis}, we investigate the underlying mechanisms of supratransmission through asymptotic analysis, deriving approximate expressions for the threshold driving amplitudes in both regimes. Section~\ref{conc1} explores the origin of rogue wave-like structures by simulating the instability of bulk solitons in an infinite-domain setting, revealing dynamics that closely resemble those previously attributed to supratransmission. In Section~\ref{conc2}, we validate our analytical predictions and numerical findings by comparing the discrete nonlinear Schr\"odinger approximation with full simulations of the original damped-driven Klein-Gordon model. Finally, we conclude with a brief discussion of potential future directions.

\section{Mathematical Model}\label{sec2}

\begin{figure}[tb]
    \centering
    \includegraphics[width=0.7\linewidth]{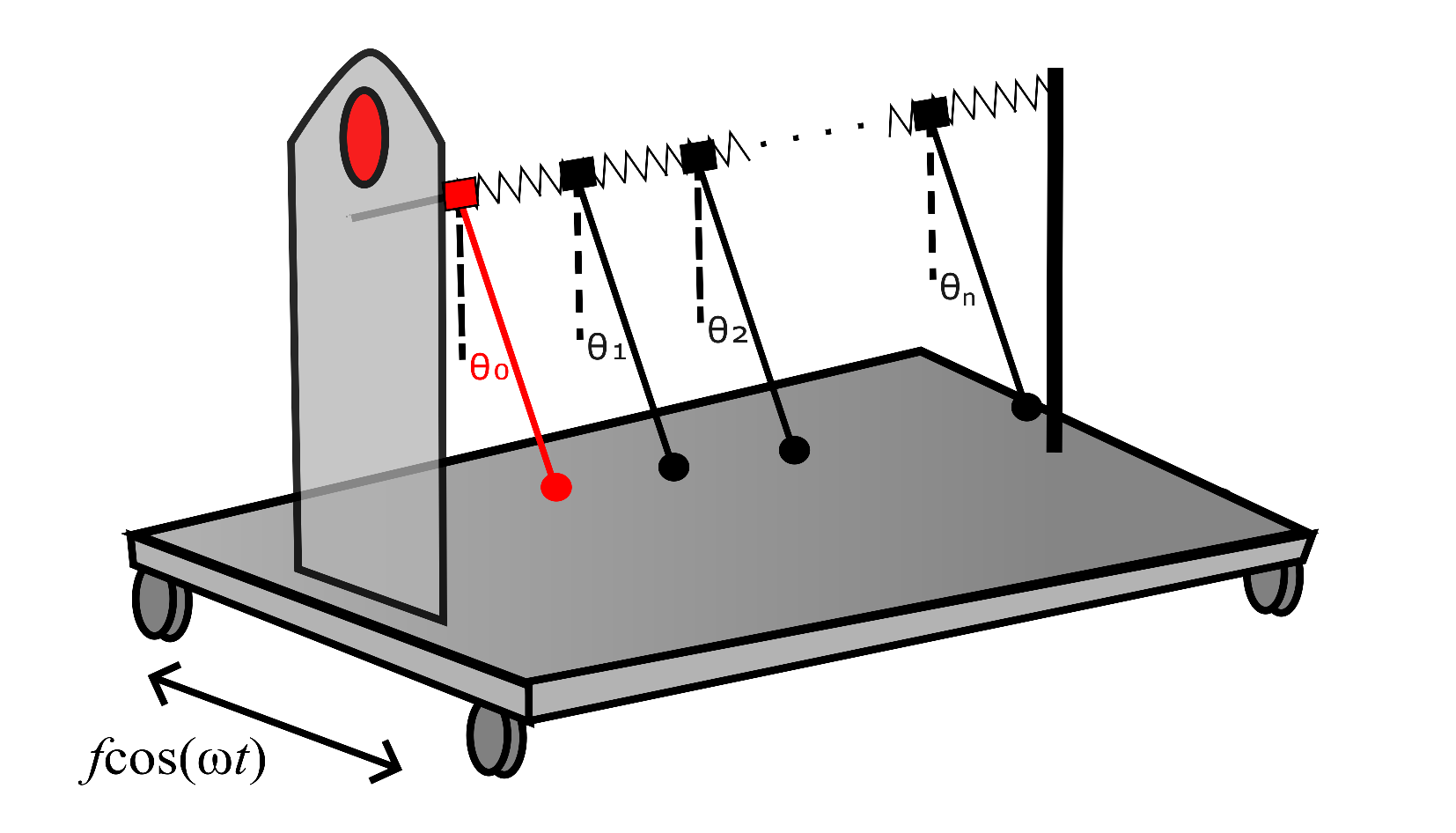}
    \caption{Illustration of an experimental pendulum chain with edge drive and external drive. }
    \label{fig:exp}
\end{figure}

The physical system considered by Motcheyo et al.\ \cite{motcheyo2022supratransmission,motcheyo2024nonlinear} consists of a chain of pendula, each connected by torsional springs. These pendula are driven by a horizontal force with a certain frequency and amplitude. Each pendulum comprises a rigid rod.
This setup forms the basis of the experimental system demonstrated in the pendulum chain \cite{thakur2007driven,english2014experimental}, as illustrated in Fig.\ \ref{fig:exp}. The dimensionless governing equation describing the system is given by \cite{cuevas2009discrete,xu2014instability}
\begin{equation} \label{eq0}
\begin{array}{l}
\ddot{\theta}_{n}=\epsilon\Delta\theta_{n}-\sin(\theta_{n})+f\omega^{2}\cos(\omega
t)\cos(\theta_{n}) -\gamma \dot{\theta}_n,
\end{array}
\end{equation}
where $\Delta\theta_{n}=\theta_{n+1}-2\theta_{n}+\theta_{n-1}$ is the discrete Laplacian representing the coupling between neighboring pendula and $\epsilon$ is the coupling strength parameter. The periodic horizontal displacement of the pivot has a frequency $\omega$ and an amplitude $f$. Here, damping from the neighboring pendula is neglected as it is considerably smaller than the onsite damping with coefficient $\gamma$, see \cite{xu2014instability}. Motcheyo et al.\ \cite{motcheyo2022supratransmission,motcheyo2024nonlinear} considered a semi-infinite lattice $n=1,2,\dots,$ and imposed a harmonic external drive at the boundary with amplitude $f_b$ and frequency $\omega_b$
\begin{equation}
    \theta_0 = f_b\cos{\omega_b t}.
    \label{b0}
\end{equation}

In this report, we consider specifically small-amplitude oscillations of the physical systems \eqref{eq0} with \eqref{b0} for analytical tractability. The linear plane wave solution of the undamped and undriven version of Eq.~\eqref{eq0} is given by $\theta_n \sim e^{i(kn - \omega t)}$, where the frequency $\omega$ and wavenumber $k$ are related through the dispersion relation $\omega^2 = 1 + 2\epsilon(1 - \cos k)$. This defines the linear spectral band as the interval $\omega^2 \in [1, 1 + 4\epsilon]$. In this work, we restrict our attention to admissible driving frequencies within the finite band gap $0 \leq \omega < 1$. Taking the standard rotating wave approximation,
\begin{equation}
    \theta_n(t) = 2\sqrt\epsilon u_n(T_1)e^{i(T_0-\Omega T_1)}+c.c.+\dots,
\end{equation}
where we also assume that $\omega=1-\epsilon\Omega/2$, $\gamma=\epsilon\Gamma$, $f=4\epsilon^{3/2} F$, $T_0=t$, and $T_1=\epsilon t/2$ (see, e.g., \cite{muda2020reduction}), and substituting the scaling into the governing equation \eqref{eq0}, from the leading terms of order $\mathcal{O}\left(\epsilon^{3/2}e^{i(T_0-\Omega T_1)}\right)$ we obtain the damped driven discrete nonlinear Schr\"odinger equation 
\begin{equation}
    i\dot{u}_n=\Delta u_n - \Omega u_n + 2|u_n|^2u_n-i\Gamma u_n+F,
    \label{main1}
\end{equation}
where the upper dot is the derivative with respect to the time variable $T_1$. The model \eqref{main1} is also known as the discrete Lugiato-Lefever equation \cite{lugiato1987spatial}. By scaling the drive at the boundary as $f_b = 4\sqrt{\epsilon}F_b$ and $\omega_b=1-\epsilon\Omega_b/2$, we obtain that the boundary condition \eqref{b0} becomes
\begin{equation}
    u_0 = F_b e^{i(\Omega-\Omega_b)T_1}.
    \label{main2}
\end{equation}
Our study will concentrate on Eqs.\ \eqref{main1} and \eqref{main2}. While in general, $\Omega$ can differ from $\Omega_b$, in this report, we will limit ourselves to the case when they are the same. The error made by the Eq.\ \eqref{main1} in approximating the dynamics of \eqref{eq0} may be inferred from \cite{muda2020reduction}. 

Note that the discrete nonlinear Schr\"odinger approximation is formally valid only for moderate detuning, i.e., $\Omega = \mathcal{O}(1)$. Nevertheless, we also examine the regime $\Omega \gg 1$ by treating the Schr\"odinger model as an independent system in its own right, rather than as an approximation of the original Klein-Gordon dynamics. Interestingly, we will demonstrate later that despite being outside the formal validity range, the Schr\"odinger model still captures dynamics of the full Klein-Gordon equation.

\section{Uniform solutions and stability\label{sec:uniform_sol}}

\begin{figure}[tbhp]
    \centering
    \includegraphics[width=0.9\linewidth]{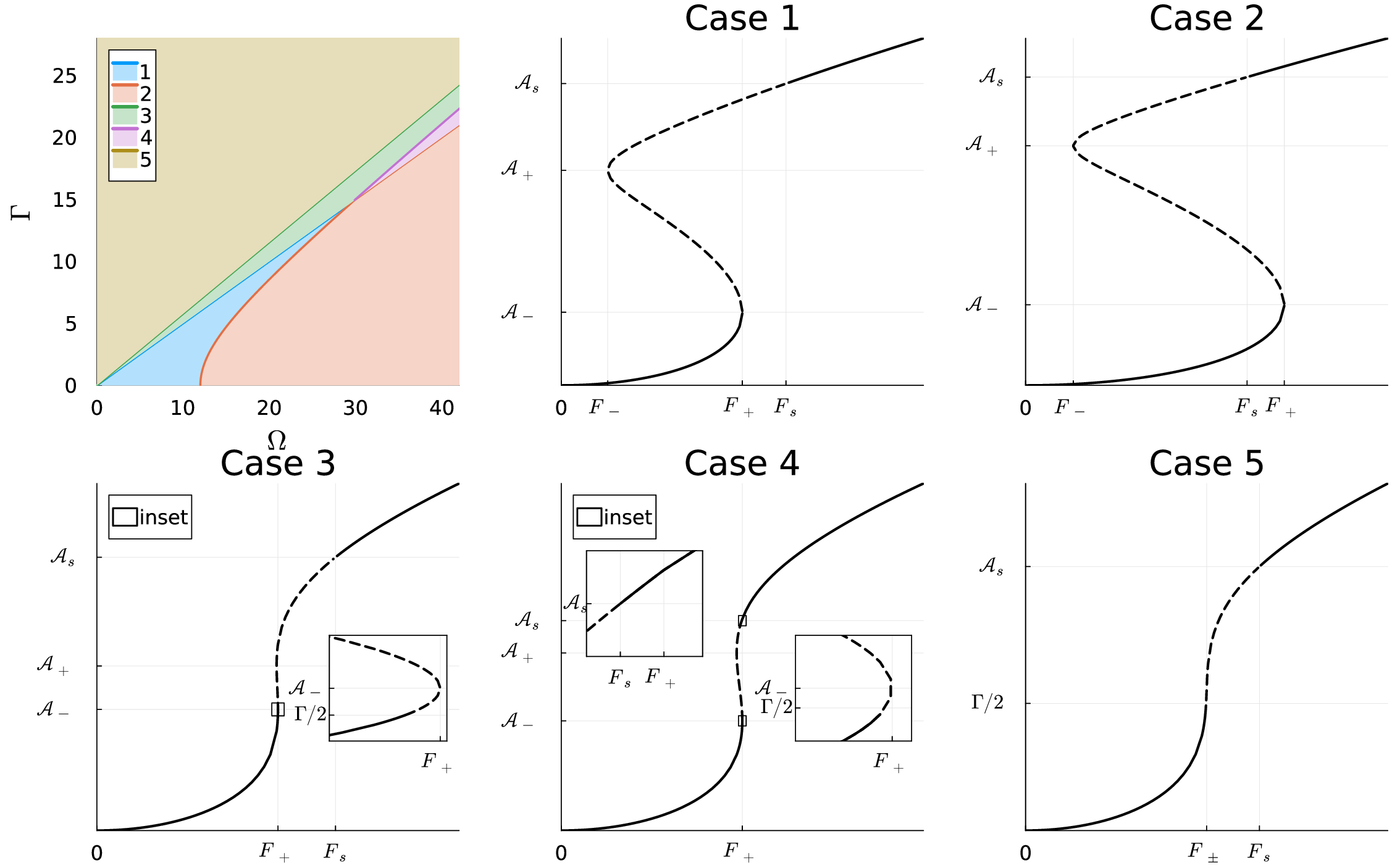}
    \caption{Stability diagram of the uniform solutions. The upper left figure shows five regions of the stability diagram in \( (\Omega, \Gamma) \) space; see Sec.\ \ref{sec:uniform_sol} for an explanation of the regions. The remaining subplots represent the stability diagram for each region, plotted in \( (F, \mathcal{A}) \) space.}
    \label{fig:uniform_stability}
\end{figure}

Due to the external drive $F$, the far-field background $\lim_{n\to\infty} u_n$ does not vanish. Therefore, it is essential to begin by analyzing the uniform solution of the model Eq.\ \eqref{main1} under the assumption of an infinite domain. Similar analyses have been conducted for the continuous counterpart, as seen in \cite{terrones1990stability, barashenkov1996existence}. Many of the stability results in our discrete model are analogous to those in the continuous case.

The uniform solutions correspond to the case where \( u_n = U \) and \( \dot{u}_n = 0 \), with \( U \) being the roots of the following function:
\begin{equation}
    P(U) := 2|U|^2 U - (i\Gamma + \Omega) U + F.
    \label{cubic_eq}
\end{equation}
Without loss of generality, we can assume that \( F \) is real. By expressing the variable in polar form as \( G e^{i\phi} = (2\mathcal{A} - \Omega) - i\Gamma \), where \( \mathcal{A} = |U|^2 \), Eq.\ \eqref{cubic_eq} becomes a linear equation with the solution:
\begin{equation}
    U = -\frac{F}{G} e^{-i\phi}.
\end{equation}
This relation holds if and only if \( \mathcal{A} \) satisfies the following cubic equation:
\begin{equation}
    \mathcal{P}(\mathcal{A}) := \left((2\mathcal{A} - \Omega)^2 + \Gamma^2\right)\mathcal{A} - F^2.
    \label{cubic2_eq}
\end{equation}
Note that Eq.\ \eqref{cubic2_eq} is now expressed in real variables, unlike the complex form in Eq.\ \eqref{cubic_eq}. This allows us to apply the formula for cubic roots from \cite{nickalls1993new}, yielding the following three roots:
\begin{equation}
    \mathcal{A}^{(j)} = \frac{\Omega}{3} \left(1 + \sqrt{1 - 3\Phi^2} \cos\left(\eta - \frac{2\pi j}{3}\right) \right),
    \label{A}
\end{equation}
for \( j = 1, 2, 3 \), where \( \Phi = \frac{\Gamma}{\Omega} \), and
\begin{equation}
    \cos(3\eta) = \frac{\frac{2F^2}{F_*^2} - (1 + 9\Phi^2)}{(1 - 3\Phi^2)^{\frac{3}{2}}}.
\end{equation}
Here, \( F_* = \sqrt{\frac{2}{27}} \Omega^{3/2} \). For each \( \mathcal{A}^{(j)} \), the corresponding \( U^{(j)} \) can be computed using the relation:
\begin{equation}
    U^{(j)} = -\frac{F}{(2\mathcal{A}^{(j)} - \Omega) - i\Gamma}.
    \label{eq:Uj}
\end{equation}

These three real roots exist only for \( F_-^2 < F^2 < F_+^2 \), where:
\begin{equation}
    F_{\pm}^2 = \frac{F_*^2}{2} \left(1 + 9\Phi^2 \pm \left(1 - 3\Phi^2\right)^{\frac{3}{2}}\right).
    \label{eq:F_turning}
\end{equation}
For the undamped case (\( \Gamma = 0 \)), \( F_+ = F_* \). If the damping exceeds the critical value \( \Gamma = \Omega / \sqrt{3} \), the system has only one uniform solution for all \( F \). We define the inflection points \( \mathcal{A}_{\pm} \) as:
\begin{equation}
    \mathcal{A}_{\pm} = \frac{\Omega}{3} \left(1 \pm \frac{1}{2} \sqrt{1 - 3\Phi^2}\right).
\end{equation}

The stability of uniform solutions for the continuous nonlinear Schr\"odinger equation with damping and external forcing has been discussed in \cite{terrones1990stability, barashenkov1996existence,diamantidis2021ExcitingExtreme}. Here, we analyze the discrete counterpart. While some results are analogous, certain stability calculations differ due to the discrete nature of the system.

We consider the perturbed solution \( u_n = U + v_n \), where \( U \) is the uniform solution. By splitting \( v_n \) into its real and imaginary parts, \( v_n = v_{n,r} + i v_{n,i} \), and setting \( v_{n,r} = c_r e^{ikn + \lambda t} \) and \( v_{n,i} = c_i e^{ikn + \lambda t} \), we obtain the following eigenvalue problem:
\begin{equation}
    \begin{bmatrix} 
        -\Gamma + 4U_r U_i & -4\sin^2(k/2) - \Omega + 2U_r^2 + 6U_i^2 \\ 
        4\sin^2(k/2) + \Omega - 6U_r^2 - 2U_i^2 & -\Gamma - 4U_r U_i 
    \end{bmatrix} 
    \begin{bmatrix} c_r \\ c_i \end{bmatrix}  
    = \lambda \begin{bmatrix} c_r \\ c_i \end{bmatrix}.
\end{equation}
The eigenvalues are given by:
\begin{equation}
    \lambda = -\Gamma \pm \sqrt{-D},
\end{equation}
where
\begin{equation}
    D(k, \Omega, \mathcal{A}) = \left(4\sin^2(k/2) + \Omega - 2\mathcal{A}\right) \left(4\sin^2(k/2) + \Omega - 6\mathcal{A}\right).
\end{equation}
The system is stable if \( D \geq -\Gamma^2 \). The minimum of \( D \) is given by:
\begin{equation}
    \min_k D = 
    \begin{cases}
        (\Omega - 2\mathcal{A})(\Omega - 6\mathcal{A}) & 0 < \mathcal{A} < \frac{\Omega}{4}, \\
        -4\mathcal{A}^2 & \frac{\Omega}{4} \leq \mathcal{A} \leq 1 + \frac{\Omega}{4}, \\
        \left(4 + \Omega - 2\mathcal{A}\right)\left(4 + \Omega - 6\mathcal{A}\right) & \mathcal{A} > 1 + \frac{\Omega}{4}.
    \end{cases}
\end{equation}

Now, consider the case where \( 0 \leq \Phi \leq \frac{1}{2} \). In this region, \( \mathcal{A}_- < \frac{\Omega}{4} \), and the stability condition is:
\begin{equation}
    (\Omega - 2\mathcal{A})(\Omega - 6\mathcal{A}) \geq -\Gamma^2,
\end{equation}
which holds for \( \mathcal{A} < \mathcal{A}_- \). Therefore, the entire lower branch is stable within this range. On the other hand, if \( \frac{1}{2} \leq \Phi \leq \frac{1}{\sqrt{3}} \), then \( \mathcal{A}_- > \frac{\Omega}{4} \), and the stability condition becomes \( \mathcal{A} > \Gamma/2 \), meaning part of the lower branch is unstable. The middle branch is always unstable. For \( \Phi > \frac{1}{\sqrt{3}} \), there is only one branch, and the solution is stable for \( \mathcal{A} < \Gamma/2 \). These results are analogous to the continuous counterpart; see \cite{barashenkov1996existence}.

A key feature of the discrete system is that, in the \( (F, \mathcal{A}) \) curve, the upper branch is initially unstable but becomes stable for large \( \mathcal{A} \). In contrast, the entire upper branch remains unstable in the continuous case. To analyze this, we define:
\begin{equation}
    \mathcal{A}_s = \frac{4 + \Omega}{3} + \frac{\sqrt{(4 + \Omega)^2 - 3\Gamma^2}}{6}, \quad 
    \mathcal{A}_{--} = \frac{\Omega}{3}\left(1 + \sqrt{1 - 3\Phi^2}\right).
\end{equation}
Here, \( \mathcal{A}_{--} \) is the other root of \( \mathcal{P} \) for \( F = F_+ \) (other than \( \mathcal{A}_- \), which has an algebraic multiplicity of two), while \( \mathcal{A}_s \) marks the boundary between the unstable and stable upper branch. Specifically, \( \mathcal{A}_+ < \mathcal{A} < \mathcal{A}_s \) is unstable, and \( \mathcal{A} > \mathcal{A}_s \) is stable. Let \( F_s \) be the corresponding drive value at \( \mathcal{A}_s \). Note that there are two possibilities for \( F_s \): either \( F_- < F_s \leq F_+ \) or \( F_s > F_+ \). To characterize this, we compare \( \mathcal{A}_s \) and \( \mathcal{A}_{--} \) by defining:
\begin{equation}
    \mathcal{S}(\Omega, \Gamma) = \sqrt{\Omega^2 - 3\Gamma^2} - 4 - \frac{\sqrt{(4 + \Omega)^2 - 3\Gamma^2}}{2}.
\end{equation}
The condition \( F_s \lessgtr F_+ \) is equivalent to \( \mathcal{S} \lessgtr 0 \). Note that \( \mathcal{S}(\Omega, \Gamma) = 0 \) has a solution only for \( \Omega \geq 12 \), and in the limit as \( \Omega \to \infty \), \( \mathcal{S} = 0 \) is satisfied when \( \Phi = \frac{1}{\sqrt{3}} \).

The five cases are summarized in different regions, as shown in the upper left of Fig.\ \ref{fig:uniform_stability}. Each case corresponds to a specific subplot in Fig.\ \ref{fig:uniform_stability}, illustrating the respective stability diagram. The five cases are as follows:
\begin{enumerate}
    \item \textbf{Case 1}: \( 0 \leq \Phi \leq \frac{1}{2} \) and \( \mathcal{S} < 0 \). The lower branch is stable for \( F_- < F < F_+ \), while the middle and upper branches are unstable. The lower branch for \( F < F_- \) and the upper branch for \( \mathcal{A} > \mathcal{A}_s \) are stable.
    \item \textbf{Case 2}: \( 0 \leq \Phi \leq \frac{1}{2} \) and \( \mathcal{S} > 0 \). The lower branch is stable for \( F_- < F < F_s \), while the middle and upper branches are unstable. In the range \( F_s < F < F_+ \), the lower and upper fixed points are stable, while the middle branch is unstable. For \( F > F_+ \), only the upper solution exists and is stable.
    \item \textbf{Case 3}: \( \frac{1}{2} \leq \Phi \leq \frac{1}{\sqrt{3}} \) and \( \mathcal{S} < 0 \). The lower branch is stable for \( F_- < F < F_+ \), while the middle and upper branches are unstable. The lower branch for \( F < F_- \) and the upper branch for \( \mathcal{A} > \mathcal{A}_s \) are stable.
    \item \textbf{Case 4}: \( \frac{1}{2} \leq \Phi \leq \frac{1}{\sqrt{3}} \) and \( \mathcal{S} > 0 \). The lower branch is stable for \( F_- < F < F_+ \), while the middle and upper branches are unstable. The lower branch for \( F < F_- \) and the upper branch for \( \mathcal{A} > \mathcal{A}_s \) are stable.
    \item \textbf{Case 5}: \( \Phi \geq \frac{1}{\sqrt{3}} \). In this case, \( \mathcal{S} < 0 \), and there is only one branch. The fixed point is stable for \( \mathcal{A} < \Gamma/2 \) or \( \mathcal{A} > \mathcal{A}_s \) and unstable otherwise.
\end{enumerate}

\section{Supratransmission and breathing solitons}\label{sec3}

\begin{figure}[tb]
    \centering
    \includegraphics[width=0.9\textwidth]{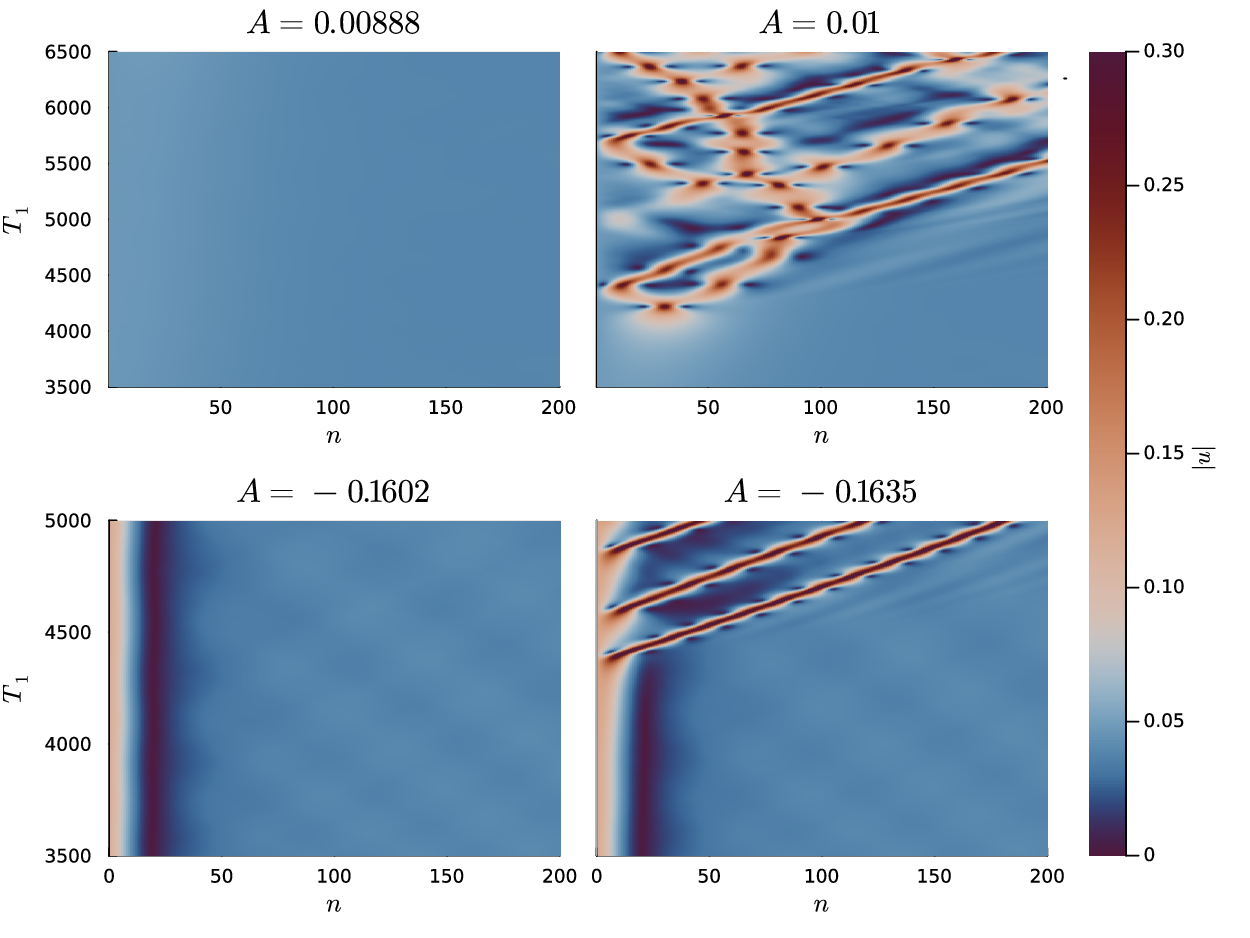}
    \caption{ Numerical simulations of the governing equations \eqref{main1} and \eqref{main2} are presented for parameter values \( \Omega = 0.01 \), \( F = 0.00027 \), and \( \Gamma = 0 \). The figures display top views of the intensity \( |u| \). The left panels correspond to a driving edge amplitude below the transmission threshold, while the right panels depict the response for an amplitude exceeding the threshold. Supratransmission, manifested as the emission of gap solitons, is observed in the right panels.}
    \label{fig:num_heatmap}
\end{figure}

In the following section, our goal is to simulate the supratransmission of a rogue wave, similar to the phenomenon observed in \cite{motcheyo2022supratransmission,motcheyo2024nonlinear}. We numerically integrate the model Eq.\ \ \eqref{main1} on a semi-infinite lattice using the boundary condition Eq.\ \eqref{main2}. The time integration is performed using a fourth-order Runge-Kutta scheme with a constant time step of \( h = 0.1 \). We focus on simulating Case 1 of the stability condition, where three uniform solutions exist.

Since numerical simulations are limited to a finite lattice, we use \( N \) spatial grid points and impose a Dirichlet boundary condition \( u_{N+1} = U^{(2)} \), where \( U^{(2)} \) represents the stable far-field background (see Eq.\ \eqref{eq:Uj}). To minimize reflections from the right computational boundary, we introduce damping by adding the term \( -i\alpha_n(u_n - U^{(2)}) \) to the right-hand side of Eq.\ \eqref{main1}. The dissipation array \( \alpha_n \) is defined as:
\begin{equation}
    \label{damp}
    \alpha_n = 
    \begin{cases} 
        0, & n = 1, \dots, N_d, \\
        \alpha_{\text{max}} \left(\frac{n - N_d}{N - N_d}\right), & n = N_d + 1, \dots, N.
    \end{cases}
\end{equation}
This setup ensures that artificial damping increases linearly from zero to \( \alpha_{\text{max}} \) over the last \( (N - N_d) \) grid points. For our simulations, we use \( \alpha_{\text{max}} = 5 \), \( N_d = 450 \), and \( N = 500 \). The system is initialized with \( u_n(0) = U^{(2)} \) for all \( n = 1, \dots, N \). To avoid an initial shock, the amplitude of the harmonic force (Eq.\ \eqref{main2}) is turned on adiabatically as:
\begin{equation}\label{eq:drive_adiabatic}
    F_b = U^{(2)} + A \left(1 - e^{-T_1 / \tau}\right),
\end{equation}
where \( \tau \gg 1 \) is the ramping time constant. In the long-time limit, \( F_b = U^{(2)} + A \). As we will demonstrate, there are upper and lower critical thresholds, \( A_c^+ \) and \( A_c^- \), respectively, for the edge drive. Crossing these thresholds results in a qualitative change in the system's dynamics. The negative sign in the edge drive \( f_b \) indicates an out-of-phase oscillation relative to the external drive \(f\) in the Klein--Gordon equation~\eqref{eq0}.

\subsection{The case of small $\Omega$\label{subsec:dyn_small_omega}}

We first focus on the case of small \( \Omega \), which provides insight into the behavior of the system when the pendulum frequency is slightly below the allowed band. Although we use different parameters, the simulated scenario is qualitatively similar to the rogue wave supratransmission discussed in \cite{motcheyo2022supratransmission,motcheyo2024nonlinear}. Specifically, we consider the case where \( \Omega = 0.01 \) and \( \Gamma = 0 \). The analysis of the undamped and undriven system, i.e., \( F = \Gamma = 0 \), was presented in~\cite{khomeriki2004nonlinear,susanto2008boundary,susanto2008calculated}. The presence of only damping still allows for supratransmission, although the resulting solitons are progressively attenuated as they propagate away from the edge. Here, we set \( F = 0.00027 \), which is close to \( F_+ = \sqrt{\frac{2}{270}} \), which is the upper limit for the existence of three uniform solutions.

Figure \ref{fig:num_heatmap} shows the heatmap of numerical simulations for various values of the driving amplitude \( A \). The panels clearly demonstrate the presence of amplitude thresholds \( A_c^\pm \), which separate two distinct dynamical regimes. The system remains in a stable evanescent state for amplitudes slightly below \( A_c^\pm \). In contrast, for amplitudes slightly larger than \( A_c^+ \) (or smaller than \( A_c^- \)), a distinct wave pattern emerges. In these simulations, the wave emergence occurs shortly after \( T_1 = 4500 \) and propagates across the lattice. This wave emergence represents the expected supratransmission, where the transmitted wave exhibits a breathing pattern. For larger deviations from the thresholds (i.e., \( A - A_c^+ \) or \( A_c^- - A \)), the onset of supratransmission occurs earlier. Our simulations reveal that the upper and lower edge amplitude thresholds for supratransmission are \( A_c^+ \approx 0.00888 \) and \( A_c^- \approx -0.1602 \). 

\begin{figure}[tbhp]
	\centering
	\includegraphics[width=0.9\textwidth]{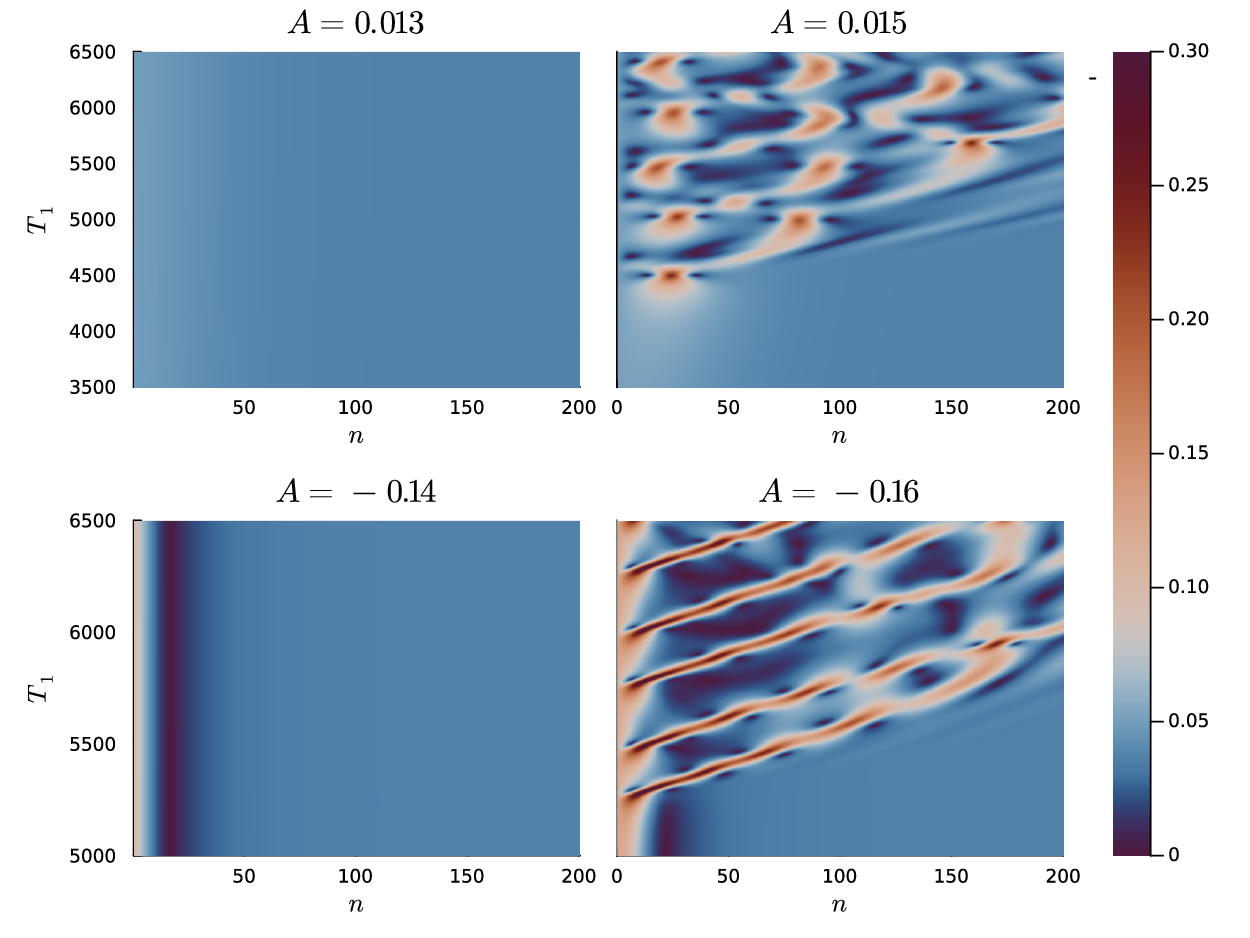}
	\caption{ Similar to Fig.\ \ref{fig:num_heatmap}  but with \(\Gamma = 0.001\).}
	\label{fig:num_heatmap_dissipation}
\end{figure}

We now consider the dynamics in the presence of damping by setting \( \Gamma > 0 \). As in the undamped case, we fix \( \Omega = 0.01 \) and \( F = 0.00027 \), but now include a damping coefficient \( \Gamma = 0.001 \). Figure~\ref{fig:num_heatmap_dissipation} shows the corresponding heatmap of numerical simulations for varying driving amplitudes \( A \). Similar to the undamped scenario, we observe clear amplitude thresholds \( A_c^\pm \) that delineate two distinct dynamical regimes. For amplitudes slightly below \( A_c^\pm \), the system relaxes to a stable evanescent state. However, for amplitudes exceeding \( A_c^+ \) (or below \( A_c^- \)), wave propagation is initiated after a delay and travels across the lattice. The presence of damping modifies the onset and structure of supratransmission: the wave amplitude is attenuated over time, and the breathing pattern observed in the undamped case becomes less pronounced. Additionally, the threshold values shift slightly due to the damping effect. Our simulations yield threshold amplitudes \( A_c^\pm \) that are larger than the undamped case.

We will discuss how these values are obtained by solving the steady-state problem described in Section \ref{sec:thr_edge_analysis} and can be approximated using the analysis provided in Subsection \ref{subsec:analysis_Omega_less1}.

\subsection{The case of large $\Omega$\label{subsec:dyn_large_omega}}

\begin{figure}[tbhp]
    \centering
    \includegraphics[width=0.6\textwidth]{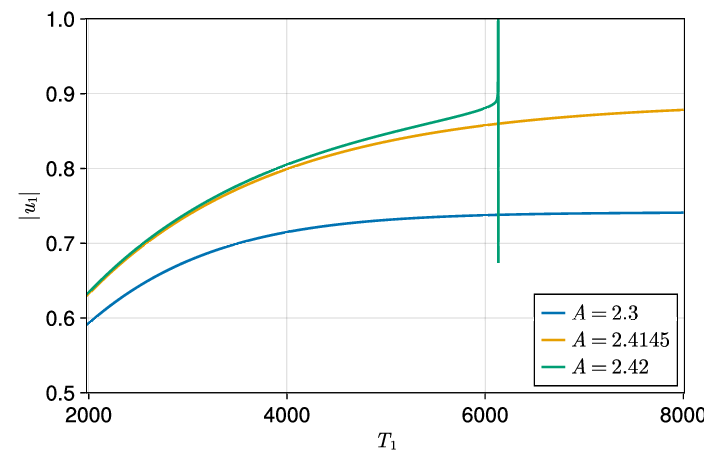}
    \caption{Time trace of \( |u_1| \) for several driving amplitudes. The parameters for the numerical simulations are \( \Omega = 3 \), \( F = 0.27 \), and \(\Gamma=0\).}
    \label{fig:num_heatmap_Om3}
\end{figure}

We conduct similar numerical simulations with the parameters \( \Omega = 3 \) and \( \Gamma = 0 \). The constant driving force is set to \( F = 0.27 \), corresponding to the case where three fixed points of the uniform solution exist. The simulations showed that \( u_1 \) has the largest modulus compared to the other lattice points. Therefore, for simplicity, only the time trace of \( u_1 \) is presented in Fig.\ \ref{fig:num_heatmap_Om3}.

In this scenario, we found that the threshold amplitude is \( A_c^+ \approx 2.4146 \). The simulation results reveal two distinct behaviors. For \( A < A_c^+ \), the variable \( u_1 \) approaches a constant value, indicating convergence to a stable fixed point. In contrast, for \( A > A_c^+ \), \( u_1 \) gradually increases over time, reaching a certain threshold before the solution undergoes a sudden and dramatic divergence. This behavior suggests an instability or the disappearance of a fixed point, leading to explosive growth in the system's dynamics. A similar behavior is observed for \( A < A_c^- \).  We omit the discussion of the damped case (\( \Gamma > 0 \)) here, as the qualitative features of the dynamics remain unchanged. In a similar effect with the case of small $\Omega$ above, the inclusion of damping only slightly shifts the threshold values and suppresses the divergence rate, without altering the overall structure of the bifurcation behavior. These dynamics will be explained by solving the steady-state problem described in Section~\ref{sec:thr_edge_analysis} and analytically derived in Subsection~\ref{subsec:analysis_Omega_great1} below.

\section{Threshold edge drive: Analysis\label{sec:thr_edge_analysis}}

This section presents an analysis of the threshold amplitudes for the edge drive, as observed in the simulations discussed earlier. The first subsection provides a numerical stability analysis of the steady-state solution to the damped-driven discrete nonlinear Schr\"odinger equation on a semi-infinite domain. This analysis is then complemented by an analytical approximation aimed at estimating the critical edge drive amplitudes \( A_c^\pm \). We consider two asymptotic regimes: the low-frequency limit \( \Omega \ll 1 \) and the high-frequency limit \( \Omega \gg 1 \). Finally, we compare the threshold amplitudes obtained numerically with those derived from the analytical approximation.

In the analysis that follows, we primarily focus on the undamped case (\( \Gamma = 0 \)) for simplicity. This choice allows for a more tractable mathematical formulation, particularly in phase plane analysis, where the system reduces to a planar two-dimensional dynamical system. Including damping would increase the dimensionality of the system and complicate the analysis without introducing qualitatively different behavior.

\subsection{Numerical methods}
\label{subsec:analysis_numerics}

To understand the dynamics presented in the previous section, we investigate the steady-state solution of Eqs.\ \eqref{main1} and \eqref{main2} on a semi-infinite domain. To achieve this, we separate Eq.\ \eqref{main1} under steady-state conditions into its real and imaginary components, subject to the boundary condition \eqref{main2} and \( u_{N+1} = U^{(2)} \) to emulate the far-field domain. We use \( N = 400 \) for this steady-state calculation. We then employ a Newton-Raphson iteration to solve the resulting nonlinear problem and obtain steady-state solutions. Following this, we analyze their stability properties. Additionally, we apply the pseudo-arclength continuation method to track solutions beyond turning points. The solutions are presented in the \( (A, \Vert u - U^{(2)} \Vert) \) space, where
\begin{equation}
    \Vert u - U^{(2)} \Vert = \left( \sum_{n=1}^N \left| u_n - U^{(2)} \right|^2 \right)^{1/2}.
\end{equation}

\begin{figure}[tbhp]
    \centering
    \subfigure[\( \Gamma = 0 \) ]{\includegraphics[width=0.47\textwidth]{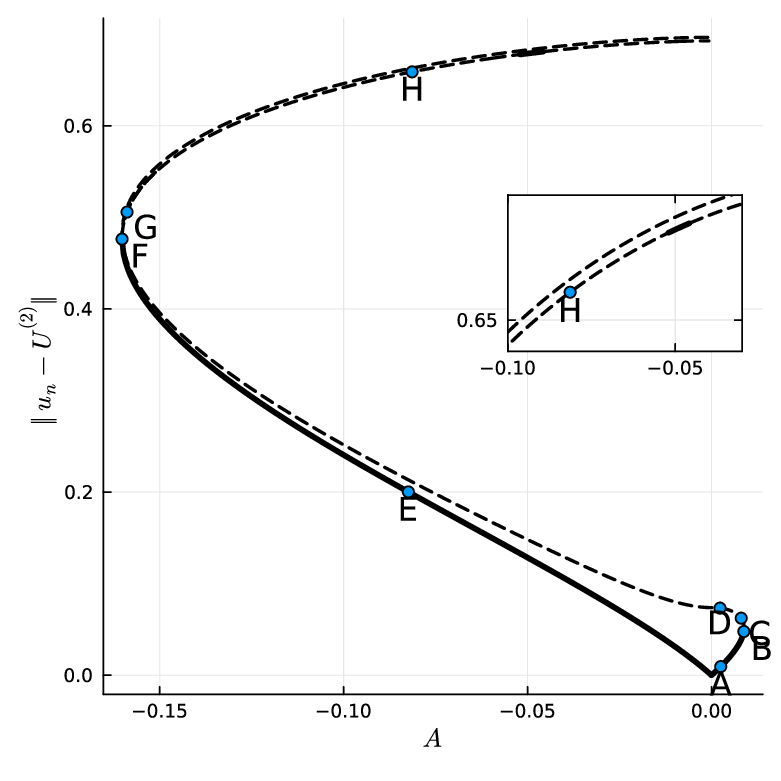}\label{fig:nbifur_diagram}}
    \subfigure[\( \Gamma = 0.001 \)] {\includegraphics[width=0.47\textwidth]{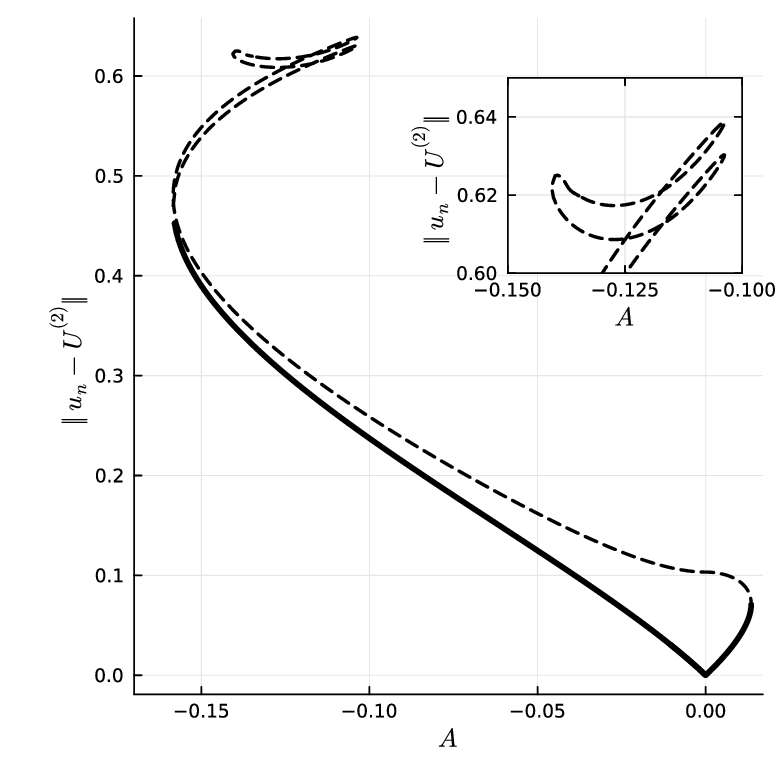}\label{fig:nbifur_diagram_dissipation}}
    \caption{Bifurcation diagrams of the evanescent wave as a function of the edge-drive amplitude for small driving frequency $\Omega$. Solid lines represent stable solutions, while dashed lines represent unstable solutions. Note that there are two stable regions on the upper branch. The parameters used for the figures are \( \Omega = 0.01 \) and \( F = 0.00027 \) with: (a) \( \Gamma = 0 \); (b) \( \Gamma = 0.001 \). 
    	The absolute profiles and the spectra at points \textsf{A}-\textsf{H} are presented in Fig.\ \ref{fig:spectrum_small}.}
    \label{fig:nbifur_diagram_combined}
\end{figure}

\begin{figure}[tbhp]
    \centering
    \includegraphics[width=0.9\textwidth]{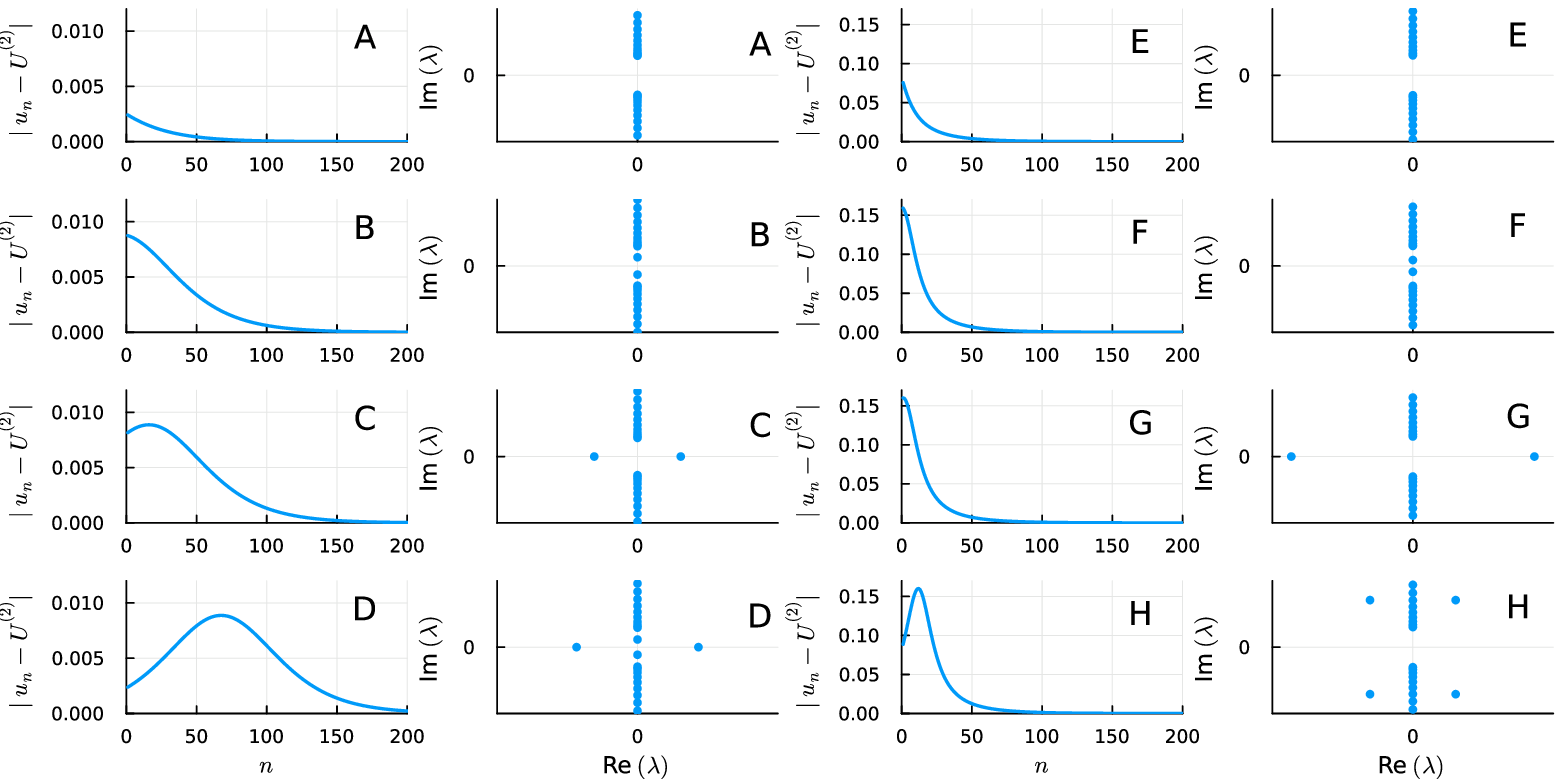}
    \caption{The first and third columns show the solution profiles at the points \textsf{A}-\textsf{H} in Fig.\ \ref{fig:nbifur_diagram}, while the second and third columns show their corresponding spectra in the complex plane. }
    \label{fig:spectrum_small}
\end{figure}

\begin{figure}[tbhp]
    \centering
    \includegraphics[width=0.6\textwidth]{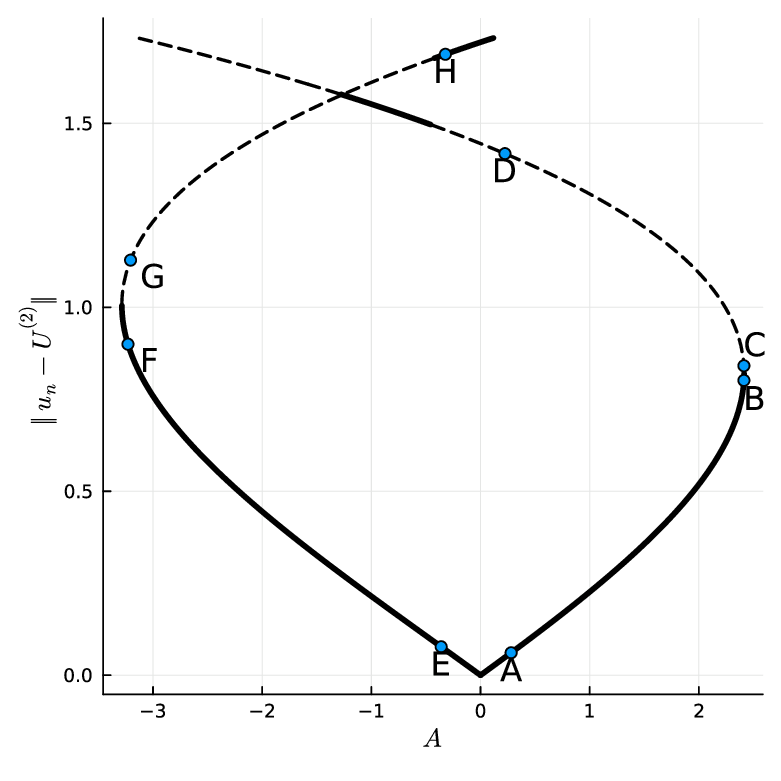}
    \caption{Similar to Fig.\ \ref{fig:nbifur_diagram_combined}, but with parameters  \( \Omega = 3 \), \( F = 0.27 \), and \( \Gamma = 0 \).  The absolute profiles and the spectra at points \textsf{A}-\textsf{H} are presented in Fig.\ \ref{fig:spectrum_diagram_om3}.}
    \label{fig:nbifur_diagram_om3}
\end{figure}

\begin{figure}[tbhp]
    \centering
    \includegraphics[width=0.9\textwidth]{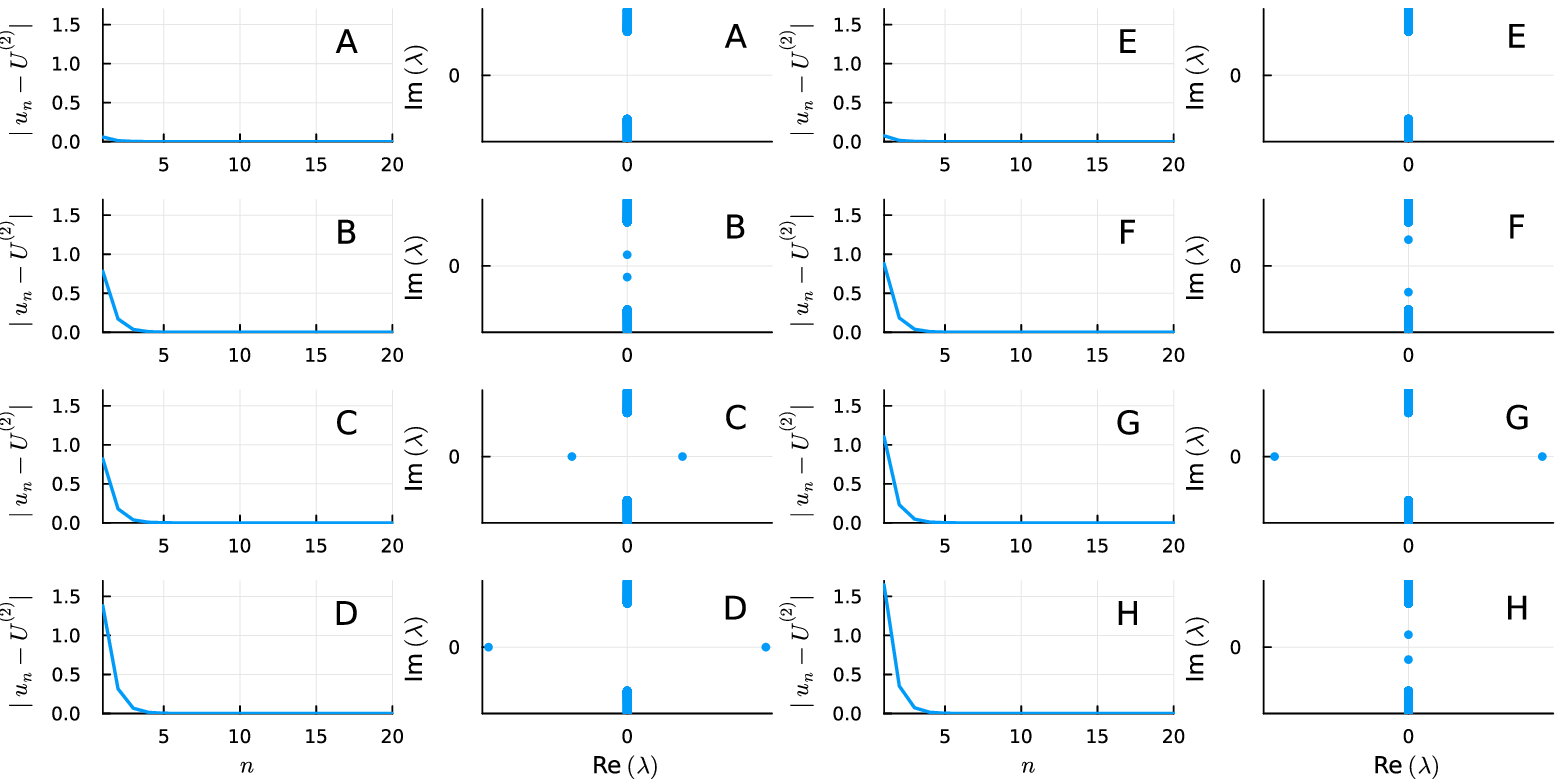}
    \caption{Similar to Fig.\ \ref{fig:spectrum_small}, but for the bifurcation diagram in Fig.\ \ref{fig:nbifur_diagram_om3}.}
    \label{fig:spectrum_diagram_om3}
\end{figure}

Once a standing wave solution is obtained, we determine its stability by computing the spectrum of its corresponding linear eigenvalue problem. To analyze the stability of these solutions, we introduce a perturbed solution of the form:
\[
u_n(t) = \tilde{u}_n + \left( r_n + is_n \right)e^{\lambda t} + \left( r_n^* + is_n^* \right)e^{\lambda^* t} ,
\]
where \(r_n \) and \(s_n\) represent small perturbations around the stationary solution \(\tilde{u}_n\), and \(\lambda\) denotes the eigenvalues associated with these perturbations. Substituting this ansatz into Eq.\ \eqref{main1} and linearizing about \(r_n\) and \(s_n\) leads to the eigenvalue problem:
\begin{equation}
	\lambda
	\begin{pmatrix}
		r_n \\ s_n
	\end{pmatrix}
	= \mathcal{J}\mathcal{L}
	\begin{pmatrix}
		r_n \\ s_n
	\end{pmatrix},
	\label{stab}
\end{equation}
where the operators \(\mathcal{J}\) and \(\mathcal{L}\) are defined as:
\[
\mathcal{J} = \begin{pmatrix}
	0_n & I_n \\
	-I_n & 0_n
\end{pmatrix}, \quad
\mathcal{L} = \begin{pmatrix}
	L_+ & M_- \\
	M_+ & L_-
\end{pmatrix}.
\]
Here \(I_n\) and \(0_n \) are identity and zero matrices with suitable size, respectively, while the operators \(L_+,L_-,M_+\) and \(M_-\) are given by:
\begin{equation}
	\begin{split}
		L_\pm &= \Delta_n-\Omega I_n +  \textrm{Re}(4|\tilde{u}_n|^2\pm2\tilde{u}_n^2) \\
        M_\pm &=\mp\Gamma I_n \pm \textrm{Im}(4|\tilde{u}_n|^2\pm2\tilde{u}_n^2),
	\end{split}
\end{equation}
where $\Delta_n$ is the discrete Laplace operator. The stability of the stationary solution \(\tilde{u}_n\) in the damped driven discrete nonlinear Schr\"odinger equation Eq.\ \eqref{main1} is determined by the spectrum of the eigenvalues \(\lambda\) obtained from Eq.\ \eqref{stab}. Specifically, a solution is linearly unstable if there exists at least one eigenvalue with \(\text{Re}(\lambda) > 0\). 

We compute the steady-state evanescent wave solutions for varying edge-drive amplitudes \( f_0 \). The corresponding bifurcation diagrams, solution profiles, and spectra are presented in Figs.~\ref{fig:nbifur_diagram_combined}--\ref{fig:spectrum_diagram_om3}, using the same parameter values as in Section~\ref{sec3}.

\subsection{The case of $\Omega\ll1$\label{subsec:analysis_Omega_less1}}

\begin{figure}[tbhp]
    \centering
    \includegraphics[width=0.8\textwidth]{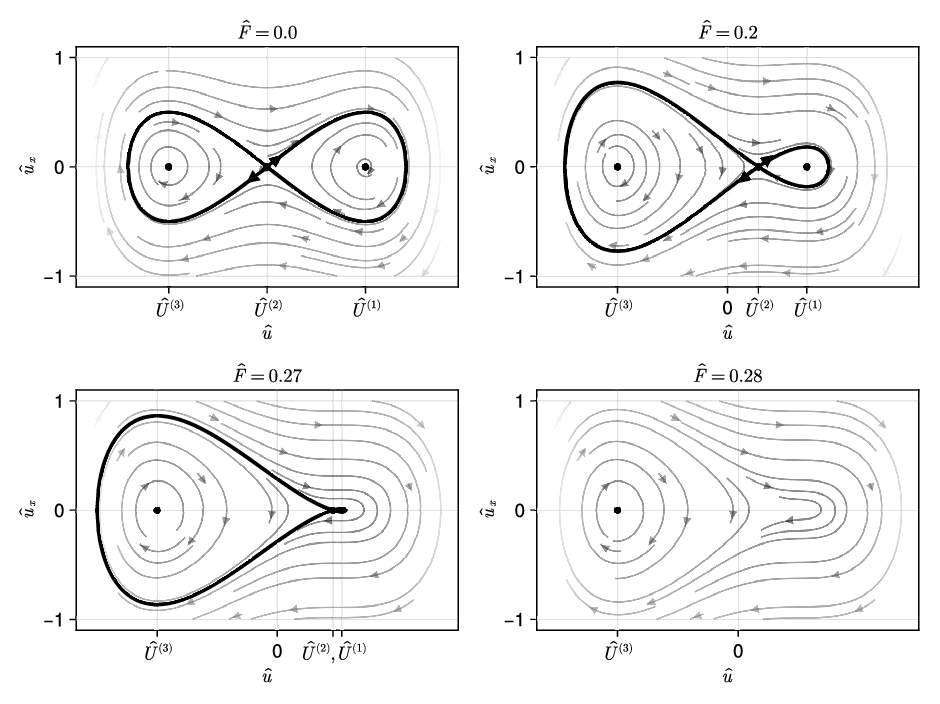}
    \caption{Phase portraits of the time-independent driven nonlinear Schr\"odinger equation, i.e., Eq.\ \eqref{main3} with \( \Gamma = 0 \) and \( \hat{u}_{\hat{T}_1} = 0 \), for various external drive $\hat{F}$. The fixed points \( \hat{U}^{(j)} = U^{(j)} / \sqrt{\Omega}  \) correspond to the flat solutions Eq.\ \eqref{A}. Note the asymmetry with the increment of $\hat{F}$. There is a critical drive above which there is no homoclinic connection.} 
    \label{fig:phase_plane}
\end{figure}

The bifurcation diagrams for the case \( \Omega \ll 1 \) are presented in Fig.~\ref{fig:nbifur_diagram_combined}. Several solution profiles and their corresponding spectra in the complex plane, taken at the marked points in the diagram, when there is no damping (i.e., Fig.\ \ref{fig:nbifur_diagram}) are shown in Fig.~\ref{fig:spectrum_small}. The left and right bifurcation curves are generally asymmetric, with the asymmetry becoming more pronounced as \( F \) approaches \( F_{\pm} \). Solutions along the lower branch are stable. Initially, there is no discrete spectrum apart from the continuous spectrum band, as illustrated at points \textsf{A} and \textsf{E}. As the edge-drive amplitude approaches the critical value \( A_c \), a pair of discrete eigenvalues detach from the continuous spectrum (points \textsf{B} and \textsf{F}) and eventually collide. They then move onto the real axis (points \textsf{C} and \textsf{G}), indicating the onset of exponential instability. On the right branch, this instability persists. On the left branch, however, the discrete eigenvalues return to the origin, collide again, and move along the imaginary axis (small bold interval before point \textsf{H}). Eventually, they collide with the continuous spectrum, forming a quartet of eigenvalues (point \textsf{H}).

The turning points of the two bifurcation curves correspond to the same threshold edge-drive amplitudes reported in Fig.~\ref{fig:num_heatmap}.
	
In the presence of damping, the overall structure of the bifurcation diagram remains qualitatively similar, see Fig.\ \ref{fig:nbifur_diagram_dissipation}. However, the turning points shift slightly due to the damping effect. Additionally, we observe an attachment and detachment of the solution branches, resulting in an imperfect bifurcation diagram. These changes reflect the breaking of the symmetry present in the undamped case, while preserving the main dynamical features.

 Next, we present an asymptotic analysis to derive the critical edge-drive amplitude, valid in the low-frequency regime \( \Omega \ll 1 \). By scaling \( T_1 = \hat{T}_1 / \Omega \), \( \Gamma = \Omega \hat{\Gamma} \), \( F = \Omega^{3/2} \hat{F} \), and writing \( u_n = \sqrt{\Omega} \hat{u}(x) \), where \( x = n \sqrt{\Omega} \), we find that slowly varying solutions of Eq.\ \eqref{main1} can be approximated by \( \hat{u}(x) \), which satisfies the continuous driven-damped nonlinear Schr\"odinger equation:
\begin{equation}
    i \hat{u}_{\hat{T}_1} = \hat{u}_{xx} - \hat{u} + 2 |\hat{u}|^2 \hat{u} - i \hat{\Gamma} \hat{u} + \hat{F}.
    \label{main3}
\end{equation}
Rogue wave solutions that may appear from this equation were studied in \cite{diamantidis2021ExcitingExtreme}. 

The phase portraits of the corresponding time-independent equation for \( \hat{\Gamma} = 0 \) and some \( \hat{F} > 0 \) are shown in Fig.\ \ref{fig:phase_plane}. There are clearly two homoclinic connections for \( |\hat{F}| < \sqrt{2/27} \) (see the presence of \( F_* \) in Eq.\ \eqref{A}). These trajectories correspond to localized solutions \( \hat{u}_\pm(x) \) that decay to \(  \hat{U}^{(2)} =  U^{(2)} / \sqrt{\Omega} \) as \( x \to \pm \infty \). The solutions, recently referred to as Barashenkov-Bogdan-Zhalan solitons \cite{feng2021barashenkov}, have an explicit expression given by \cite{barashenkov1990driven}:
\begin{eqnarray}
    \label{BBZ} 
    \hat{u}_{\pm}(x) &=& \psi_0 \left(1 + \varphi_{\pm}(x)\right) = \psi_0 \left(1 + \frac{2 \sinh^2 \varrho}{1 \pm \cosh \varrho \cosh(\Lambda x)}\right), \\
    \label{60} 
    \psi_0 &=& \frac{1}{\sqrt{2(1 + 2 \cosh^2 \varrho)}}, \quad \Lambda = \frac{\sqrt{2} \sinh \varrho}{\sqrt{1 + 2 \cosh^2 \varrho}},
\end{eqnarray}
where
\begin{equation}
    \label{BBZ1} 
    \hat{F} = \frac{\sqrt{2} \cosh^2 \varrho}{(1 + 2 \cosh^2 \varrho)^{3/2}},
\end{equation}
with \( \hat{F} \in (0, 2 / (3 \sqrt{6})) \) and \( \varrho \in (0, \infty) \). It is immediate that \( \psi_0 = \hat{U}^{(2)}  \). When \( \hat{F} \to 0 \), the soliton expression \eqref{BBZ} simplifies to:
\begin{equation}
    \hat{u}_{\pm} \to \pm \sqrt{\Omega} \, \text{sech}(\sqrt{\Omega} x).
\end{equation}
The solution \( \hat{u}_+ \) is unstable \cite{barashenkov1990driven}, while \( \hat{u}_- \) is stable for small enough \( \hat{F} \) \cite{feng2021barashenkov}. Both stability results are for \( \hat{\Gamma} = 0 \). The existence and stability of these solitons for \( \hat{\Gamma} \neq 0 \) were first studied numerically in \cite{barashenkov1996existence} and only recently analyzed analytically in \cite{bengel2023stability}. 

The evanescent standing waves observed along the bifurcation diagram in Fig.~\ref{fig:nbifur_diagram} are made from part of the localized nonlinear wave structures above. These homoclinic solutions are critical in determining the threshold amplitudes for edge-driven supratransmission. Specifically, the threshold edge-drive amplitudes correspond to the amplitudes of these localized waves, which we can precisely calculate using the relationship:
\begin{equation}
    \label{eq:Ac_small}
    A_c = \frac{2 \Omega^{3/2} \sinh^2 \varrho}{\left(1 \pm \cosh \varrho\right) \sqrt{2(1 + 2 \cosh^2 \varrho)}},
\end{equation}
where \( \varrho \) is a parameter that satisfies Eq.~\eqref{BBZ1}. 

\subsection{The case of $\Omega\gg1$\label{subsec:analysis_Omega_great1}}

The bifurcation diagrams for the case \( \Omega \gg 1 \) are shown in Fig.~\ref{fig:nbifur_diagram_om3}. Several solution profiles and their corresponding spectra in the complex plane, taken at the indicated points, are presented in Fig.~\ref{fig:spectrum_diagram_om3}. The general behavior near the turning point resembles that observed for \( \Omega \ll 1 \). However, unlike the low-frequency case, no formation of quadruplet eigenvalues is observed. Additionally, a stable region emerges as \( A \to 0^- \) on the upper left branch. The turning points also correspond to the same threshold edge-drive amplitude reported in Fig.~\ref{fig:num_heatmap_Om3}.

\begin{figure}[tbhp!]
    \centering
    \includegraphics[width = 0.9\textwidth]{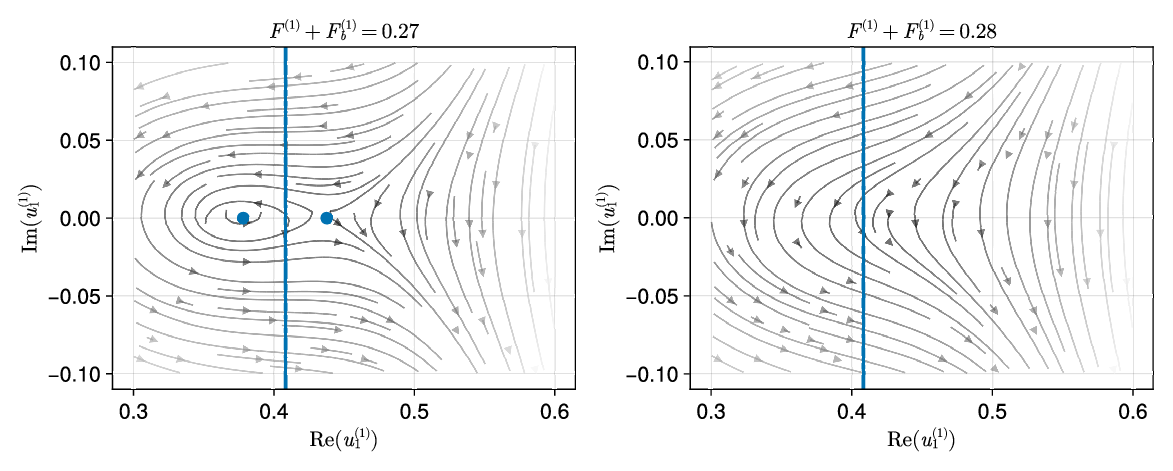}
    \caption{
        Phase portraits of the asymptotic equation for the case $\Omega \gg 1$, given by Eq.\ \eqref{eq:model_asympt_O_3/2}. The two dots in the left figure represent fixed points, which exist only when $F_b^{(1)} + F^{(1)} < \sqrt{\frac{2}{27}}$. As $F_b^{(1)} + F^{(1)} \to \sqrt{\frac{2}{27}}$, the fixed points approach each other and eventually collide at the blue line described by Eq.\ \eqref{eq:crit_u1}. For $F_b^{(1)} + F^{(1)} > \sqrt{\frac{2}{27}}$, the fixed points vanish, as shown in the right figure. The phase plane corresponds to the parameters $\Gamma = 0$. 
    }
    \label{fig:Phase_Plane_Om3}
\end{figure}

We follow the asymptotic analysis developed in~\cite{susanto2008boundary} to approximate the turning points. The variables and parameters are expanded as follows:
\[
u_n = \Omega^{1/2} u_n^{(1)} + \Omega^{-1/2} u_n^{(2)} + \dots, \quad F_b = \Omega^{3/2} F_b^{(1)} + \Omega^{1/2} F_b^{(2)} + \dots, 
\]
\[
\Gamma = \Omega \Gamma^{(1)}, \quad F = \Omega^{3/2} F^{(1)} + \Omega^{1/2} F^{(2)} + \dots, \quad T_1 = \hat{T}_1 / \Omega.
\]
From Eq.\ \eqref{main1}, we derive the equation at $\mathcal{O}(\Omega^{3/2})$:
\begin{equation}
i \frac{du_1^{(1)}}{d\hat{T}_1} = 2 |u_1^{(1)}|^2 u_1^{(1)} - (1 + i \Gamma^{(1)}) u_1^{(1)} + F_b^{(1)} + F^{(1)}. 
\label{eq:model_asympt_O_3/2}
\end{equation}
This equation describes the small amplitude dynamics of the $u_1$. Its phase planes are shown in Fig.\ \ref{fig:Phase_Plane_Om3} for two cases: $F_b^{(1)} + F^{(1)} \lessgtr \sqrt{\frac{2}{27}}$. 

In the numerical simulations, we fix $F$ (which corresponds to $F^{(1)}$) and $F_b^{(1)}$ (that corresponds to $U^{(2)} + A$). Therefore, the condition
aligns with the case $A \lessgtr A_c^+$, as discussed in Subsection \ref{subsec:dyn_large_omega}. 

For the first case ($F_b^{(1)} + F^{(1)} < \sqrt{\frac{2}{27}}$), there are two fixed points (marked as blue dots in Fig.\ \ref{fig:Phase_Plane_Om3}). The left fixed point is stable, while the right one is unstable. This explains the convergence of $u_1$ to a fixed point observed in the dynamics for $A < A_c^+$ (see Subsection \ref{subsec:dyn_large_omega}). The proof of (in)stability follows a similar approach to the stability analysis of uniform solutions in Section \ref{sec:uniform_sol}. When $F_b^{(1)} + F^{(1)} = \sqrt{\frac{2}{27}}$, the two fixed points collide (indicated by the blue line in Fig.\ \ref{fig:Phase_Plane_Om3}). Beyond this critical value, the fixed points disappear, as shown in the right plot of Fig.\ \ref{fig:Phase_Plane_Om3}. The absence of (stable) fixed points in this regime explains the blow-up behavior observed in Subsection \ref{subsec:dyn_large_omega}.

The time-independent form of Eq.\ \eqref{eq:model_asympt_O_3/2} is again a cubic equation of the form \eqref{cubic_eq}. The turning points for two of the roots can be derived from Eq.\ \eqref{eq:F_turning} as:
\begin{equation}
F_{b,c}^{(1)} = -F^{(1)} \pm \sqrt{\frac{1}{27} \left(1 + 9 {\Gamma^{(1)}}^2 + \left(1 - 3 {\Gamma^{(1)}}^2\right)^{\frac{3}{2}}\right)}.
\label{Ac1}
\end{equation}
At this critical value, the corresponding $u_1^{(1)}$ is given by:
\begin{equation}
u_{1,c}^{(1)} = \pm \frac{\sqrt{\frac{1}{3} \left(1 + 9 {\Gamma^{(1)}}^2 + \left(1 - 3 {\Gamma^{(1)}}^2\right)^{\frac{3}{2}}\right)}}{2 \left(1 + 3 {\Gamma^{(1)}}^2 + \sqrt{1 - 3 {\Gamma^{(1)}}^2}\right)} \left(1 + \sqrt{1 - 3 {\Gamma^{(1)}}^2} - 3i \Gamma^{(1)}\right).
\label{eq:crit_u1}
\end{equation}
For small $\Gamma^{(1)}$, the Taylor expansion of $u_{1,c}^{(1)}$ is:
\begin{equation}
u_{1,c}^{(1)} = \pm \left( \frac{1}{\sqrt{6}} - \frac{1}{8} \sqrt{\frac{3}{2}} {\Gamma^{(1)}}^2 - \sqrt{\frac{3}{8}} i \Gamma^{(1)} \right).
\end{equation}
From the equation at $\mathcal{O}(\Omega^{1/2})$, we obtain the next correction to the critical drive, which is:
\begin{equation}
F_{b,c}^{(2)} = -F^{(2)} + 2 u_{1,c}^{(1)}.
\end{equation}
One can continue the calculation to obtain a higher-order approximation. Based on the above analysis, the turning point can be approximated as: 
\begin{equation}\label{eq:Ac_large}
    A_c^\pm = -F-U^{(2)}\pm \left(\Omega^{3/2}\sqrt{\frac{1}{27}\left(1+9{\Gamma^{(1)}}^2+ \left(1-3{\Gamma^{(1)}}^2\right)^{\frac{3}{2}}\right)} + 2\Omega^{1/2} u_{1,c}^{(1)}\right) . 
\end{equation}
The spectrum of the fixed points can also be calculated following \cite{susanto2008boundary,peschel2004discrete,egorov2013spontaneously,johansson2019stability}. However, we leave it to the interested reader.

\subsection{Threshold edge-drive amplitudes}

\begin{figure}[tbhp!]
    \centering
    \includegraphics[width = 0.5\textwidth]{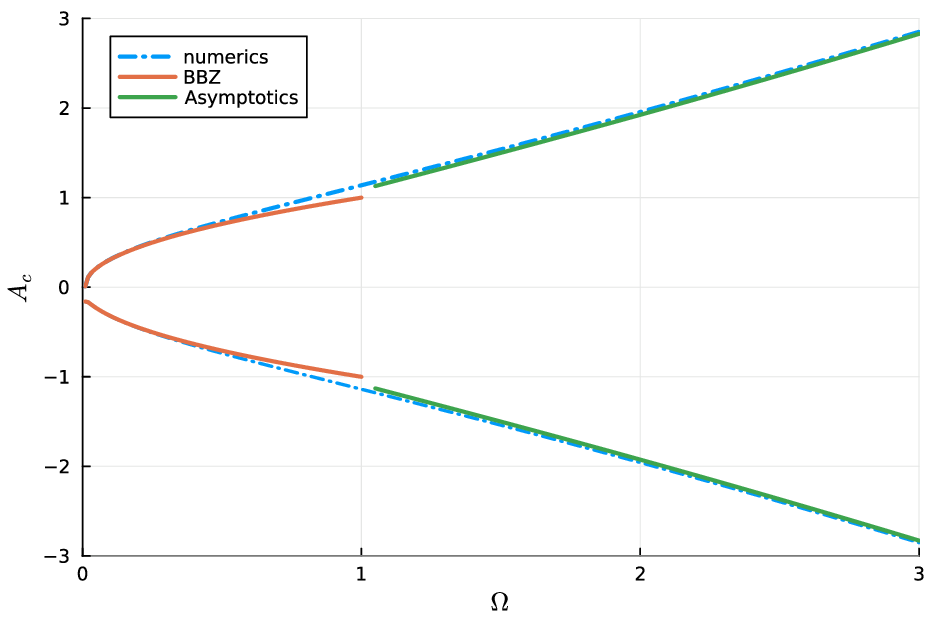}
    \caption{
        The upper and lower amplitude thresholds as functions of $\Omega$. The cyan dashed-dot line numerically represents the turning points obtained using the procedure described in Subsection \ref{subsec:analysis_numerics}. The orange line corresponds to the turning points derived from the Barashenkov-Bogdan-Zhalan solitons, i.e., $\hat{u}_{\pm}(0)$ in Eq.\ \eqref{60}. The green line shows the turning points calculated using the asymptotic method \eqref{eq:Ac_large}. All results correspond to the parameters $F = 0.00027$ and $\Gamma = 0$.
    }
    \label{fig:Om_Ac}
\end{figure}

We compute the steady-state solutions of the damped driven discrete nonlinear Schr\"odinger equation, as described in Subsection \ref{subsec:analysis_numerics}, for $\Omega \in [0.01, 3]$. Our goal is to identify the critical edge thresholds $A_c^\pm$.  The calculations use the parameters $F = 0.00027$ and $\Gamma = 0$. The numerical results are compared with the analytical predictions for the turning points derived in Subsections \ref{subsec:analysis_Omega_less1} and \ref{subsec:analysis_Omega_great1}. This comparison is presented in Fig.\ \ref{fig:Om_Ac}. As expected, the Barashenkov-Bogdanov-Zhalan soliton model \eqref{eq:Ac_small} shows good agreement with the numerical simulations for $\Omega \ll 1$, although it slightly underestimates the critical edge drive. This deviation becomes more noticeable as $\Omega \to 1$. The approximation \eqref{eq:Ac_large} also agrees well with the numerical results $A_c^\pm$ for $\Omega \gg 1$. This demonstrates the overall accuracy of the turning point calculations derived from both analytical approximations.

\section{Rogue waves on bulk solitons}
\label{conc1}

\begin{figure}[tbhp]
	\centering
	\includegraphics[width = 0.8\textwidth]{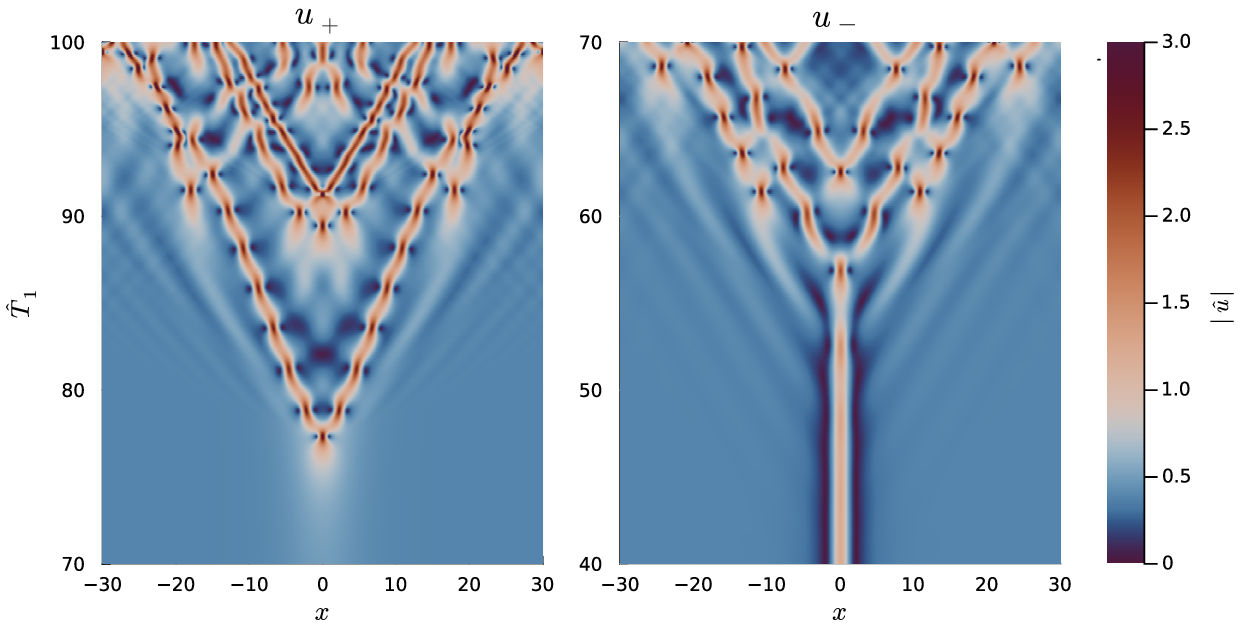}
	\caption{
		Instability of bulk solitons leading to rogue wave-like dynamics, governed by Eq.~\eqref{main3} with \( \hat{F} = 0.27 \), \( \Gamma = 0 \), and initial conditions \( \hat{u}_\pm \) as defined in Eq.~\eqref{BBZ}. The value of \( \hat{F} \) corresponds to the bottom-left panel of Fig.~\ref{fig:phase_plane}. Shown is the top view of the intensity \( |\hat{u}| \).
		}
	\label{fig:RW}
\end{figure}

We have shown that the supratransmission phenomenon that was thought to release rogue waves before \cite{motcheyo2022supratransmission,motcheyo2024nonlinear} should be attributed to the instability of the Barashenkov-Bogdan-Zhalan solitons, as discussed in Section \ref{subsec:dyn_small_omega}. We simulate the bulk solitons in an infinite one-dimensional domain to further demonstrate this. To approximate the infinite domain, we impose far-field boundary conditions \( u_0 = u_{N+1} = U^{(2)} \), with the spatial domain defined as \([-30, 30]\) and discretized using \( N = 601 \) grid points. This configuration ensures that \( u(\pm 30) \approx U^{(2)} \), effectively mimicking an infinite domain. The simulations are conducted using Eq.\ \eqref{main3} with the parameters \(\hat{F} = 0.27\) and \(\hat{\Gamma} = 0\). The initial conditions \(\hat{u}_\pm\) are derived from Eq.\ \eqref{BBZ}. These parameter values are consistent with those used in Subsection \ref{subsec:dyn_small_omega}. For numerical integration, we employ a fourth-order Runge-Kutta scheme with a fixed time step of \( d\hat{T}_1 = 0.005 \).

The results of the simulations are depicted in Fig.\ \ref{fig:RW}. The left panel illustrates the dynamical simulation starting with the initial condition \( u_+ \). The soliton \( u_+ \) is unstable \cite{barashenkov1990driven}. We have confirmed its instability by numerically solving the associated eigenvalue problem, where we obtained eigenvalues with a nonvanishing real part (this case corresponds to the limit solution along the right upper branch of Fig.\ \ref{fig:nbifur_diagram} as $A\to0^+$). We observe the abrupt emergence of a rogue wave-like structure at \(\hat{T}_1 \approx 77\). This soliton subsequently splits and propagates in both directions, exhibiting a breathing pattern throughout the simulation. The solitons propagating after the split resemble the characteristics and dynamics of \( u_- \), which will be discussed below. The dynamics of this unstable soliton are strikingly similar to those shown in the upper right corner of Fig.\ \ref{fig:num_heatmap}. 

For the initial condition \( u_- \), the soliton remains stable for small values of \(\hat{F}\) but becomes unstable as \(\hat{F}\) increases. For the specific parameters used in this study, we have confirmed the instability of \( u_- \) by numerically solving the associated eigenvalue problem, where we obtained quadruplets of complex eigenvalues (this case corresponds to the limit solution along the left upper branch of Fig.\ \ref{fig:nbifur_diagram} as $A\to0^-$). The numerical simulation further validates this instability, as shown in the right panel of Fig.\ \ref{fig:RW}. The resulting unstable soliton propagates through the domain, displaying dynamics closely resembling the supratransmission phenomenon observed in the lower right corner of Fig.\ \ref{fig:num_heatmap}.

\section{Validation with the Klein-Gordon Model} \label{conc2}

\begin{figure}[tbhp]
	\centering
	\includegraphics[width=0.7\textwidth]{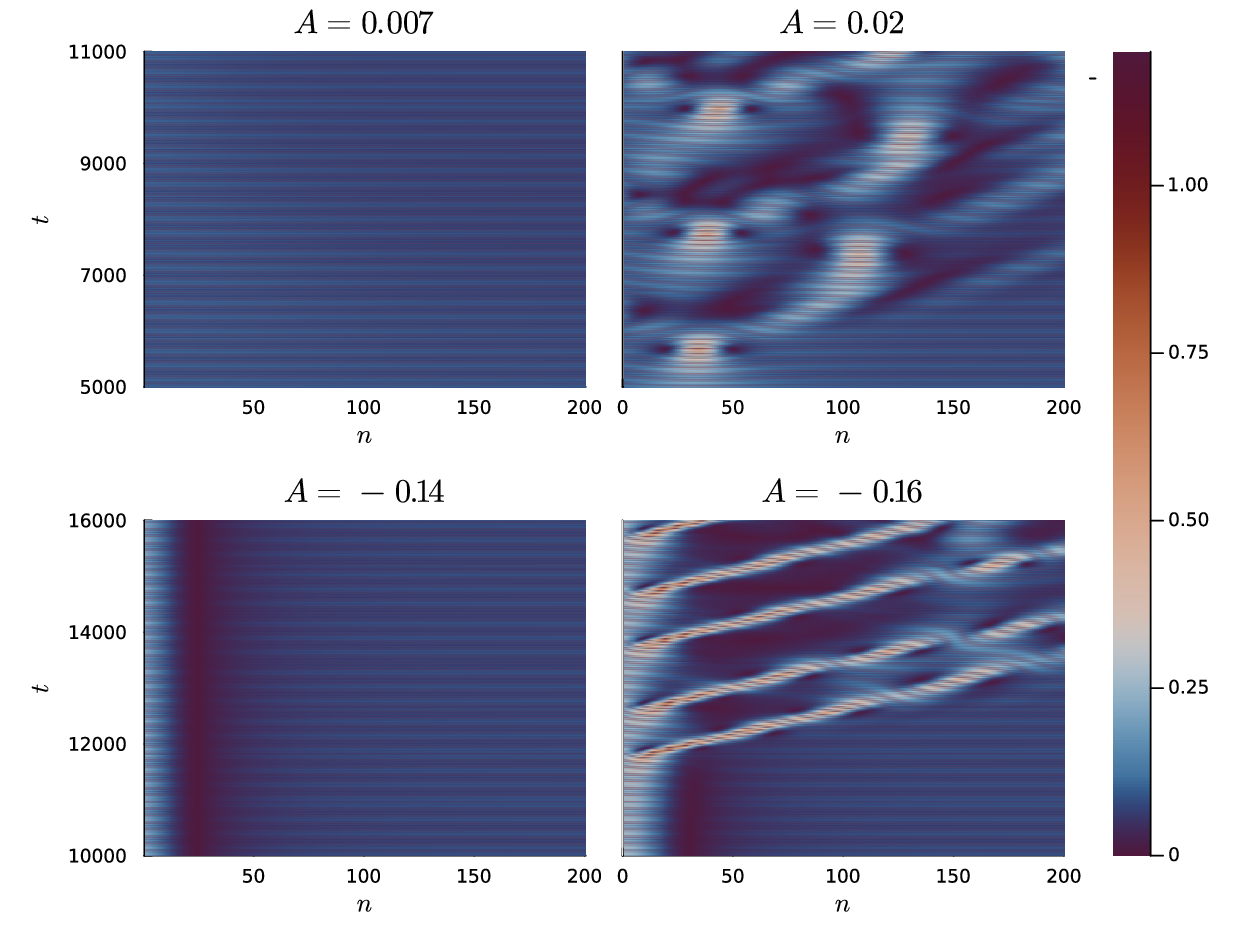}
	\caption{
		Numerical simulations of supratransmission in the damped-driven Klein-Gordon equation \eqref{eq0}. The parameter values used are $\epsilon = 0.5$, $\omega = 0.9975$, $\omega_b = 0.9975$, $\gamma = 0.0005$, and $f = 0.0003818$, which directly correspond to those used in the discrete nonlinear Schr\"odinger equation simulation shown in Fig.~\ref{fig:num_heatmap_dissipation}. The figure presents a top-view map of the spatial-temporal intensity distribution $|\theta_n|$.
	}
	\label{fig:RW_KG}
\end{figure}

The damped-driven discrete nonlinear Schr\"odinger equation \eqref{main1}, rigorously analyzed in the preceding sections, serves as a reduced model for the more general damped-driven Klein-Gordon system described by Eq.~\eqref{eq0}. To assess the validity of the analytical and numerical results obtained within the dNLS framework, we conducted numerical simulations of the full KG model using parameter values consistent with those discussed in Subsection~\ref{subsec:dyn_small_omega}. Specifically, we used $\epsilon = 0.5$, $\omega = 0.9975$, $\omega_b = 0.9975$, $\gamma = 0.0005$, and $f = 0.0003818$. Numerical integration was performed using a fourth-order Runge-Kutta scheme with a time step $h = 0.2$ over a spatial lattice comprising $N = 500$ sites. The number of sites is taken to minimize reflections during the simulation window. Neumann boundary conditions were imposed in the far field. 

In the absence of edge driving, the system naturally evolves with frequency $\omega$ and amplitude $\Theta$, which can be approximated by the stable uniform solution of the Schr\"odinger equation as $\Theta \approx 4U^{(2)}$. However, our simulations reveal that the approximation tends to slightly overestimate the steady-state amplitude observed in the full Klein-Gordon model. This discrepancy is expected due to higher-order effects neglected in the asymptotic reduction. For simulations with edge driving, the driving amplitude was modulated according to the adiabatic ramping profile given by
\begin{equation}
	f_b(t) = 4|U^{(2)}| + 4A\left(1 - e^{-t/\hat{\tau}}\right),
\end{equation}
where we selected a characteristic ramping timescale of $\hat{\tau} = 2000$. To ensure consistency between the initial state and the applied boundary conditions, the system was initialized with $\theta_n(0) = 4U^{(2)}$ and $\dot{\theta}_n(0) = 0$.

As illustrated in Fig.~\ref{fig:RW_KG}, the numerical results from the full Klein-Gordon model exhibit supratransmission features closely resembling those obtained from the nonlinear Schr\"odinger simulations (see Fig.~\ref{fig:num_heatmap_dissipation}). Despite the simplifications inherent in the reduced model, the observed amplitude thresholds in the Klein-Gordon simulations align closely with the theoretical predictions derived in Sections~\ref{subsec:analysis_numerics} and \ref{subsec:analysis_Omega_less1}. This agreement supports the validity and predictive accuracy of the Schr\"odinger approximation for capturing key dynamical features of the original lattice system.

\begin{figure}[tbhp]
	\centering
	\includegraphics[width=0.7\textwidth]{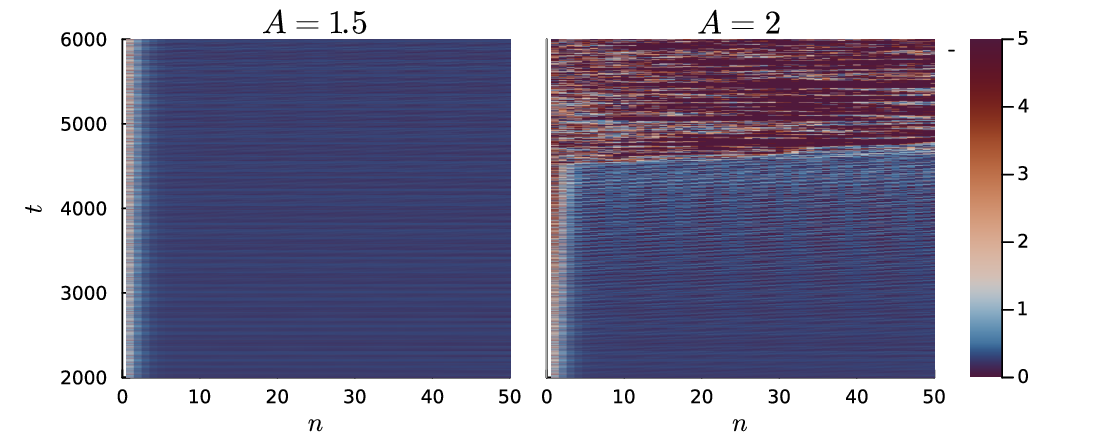}
	\caption{
		Similar to Fig.~\ref{fig:RW_KG}, but with  parameter values $\epsilon = 0.5$, $\omega = 0.25$, $\omega_b = 0.25$, $\gamma = 0.$, and $f =  0.381838$. The parameters directly correspond to those used in the discrete nonlinear Schr\"odinger equation simulation shown in Fig.~\ref{fig:num_heatmap_Om3}. 
	}
	\label{fig:RW_KG_small_om}
\end{figure}

To complement our analysis, we performed analogous simulations for the large $\Omega$. The parameters are directly correspond to Subsection~\ref{subsec:dyn_large_omega}, using $\epsilon = 0.5$, $\omega = 0.25$, $\omega_b = 0.25$, $\gamma = 0.$, and $f =  0.381838$.  As shown in Fig.~\ref{fig:RW_KG_small_om}, the resulting dynamics exhibit qualitatively similar features as the dNLS simulation discussed in  Subsection~\ref{subsec:dyn_large_omega}. However, we observe a significant quantitative discrepancy: the empirical critical amplitude $A_c^+$ is substantially  smaller than the dNLS estimate. This deviation is expected, as the the dNLS  approximation becomes less accurate of Klein-Gordon equation for large detuning.

\begin{figure}[tbhp]
	\centering
	\includegraphics[width = 0.8\textwidth]{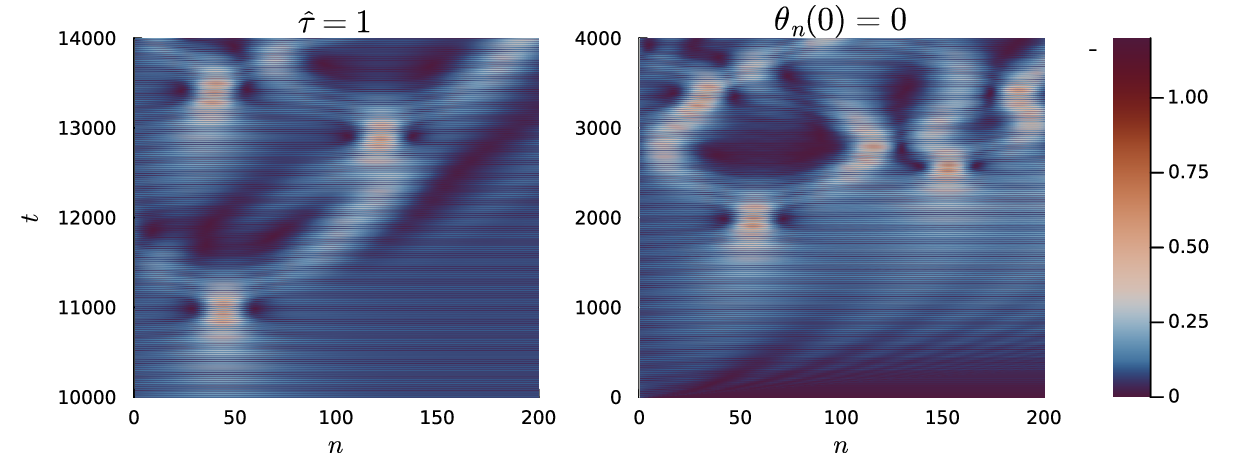}
	\caption{
				The same with the top left panel of Fig.\ \ref{fig:RW}, but with a different ramping factor $\hat{\tau}$ (left) or initial condition $\theta_{n}(0)$ (right).	}
	\label{fig:KG_IC_fail}
\end{figure}

The accuracy of supratransmission simulations in the Klein-Gordon model critically depends on the appropriate choice of initial conditions and driving protocols. Figure~\ref{fig:KG_IC_fail} highlights potential numerical artifacts that can arise from unsuitable simulation setups. In the left panel, we apply a sudden increase in the boundary driving using parameters $A = 0.007$ and $\hat{\tau} = 1$. This abrupt transition induces a spurious rogue wave centered around $n \approx 50$ at time $t \approx 11000$. Notably, this artifact is absent when the ramping timescale is sufficiently large, as demonstrated in the upper left panel of Fig.~\ref{fig:RW_KG}. Similarly, initializing the system with $\theta_n(0) = \dot{\theta}_n(0) = 0$, even in the presence of a slow ramping ($\hat{\tau} \gg 1$), can introduce discontinuities between the initial state and boundary drive. This mismatch results in artificial wave packets resembling rogue waves, as shown in the right panel of Fig.~\ref{fig:KG_IC_fail}.

Given the sensitivity of the model to initial and boundary conditions, it is essential to interpret numerically observed rogue wave phenomena with caution. In particular, the emergence of similar structures in prior work, such as those depicted in \cite[Fig.~8(a)--(b)]{motcheyo2022supratransmission}, may potentially be attributed to numerical artifacts rather than genuine physical supratransmission. 

Building upon the analytical and numerical results presented in this study, several avenues for future research may be pursued. One promising direction involves 
exploring higher-dimensional analogs of the system that will reveal richer dynamical behaviors, including transverse instabilities and two-dimensional localized structures. Furthermore, 
experimental validation of the predicted supratransmission phenomena under controlled initial conditions would provide crucial insight into the physical realizability and robustness of these effects.

\section*{Acknowledgment}
We acknowledge the contribution of Khalifa University's high-performance computing and research computing facilities in providing computational resources for this research. HS also acknowledged support by Khalifa University through a Competitive Internal Research Awards Grant (No.\ 8474000413/CIRA-2021-065) and Research \& Innovation Grants (No.\ 8474000617/RIG-S-2023-031 and No.\ 8474000789/RIG-S-2024-070).

\section*{Conflict of interest}
The authors declare that they have no conflict of interest.

\section*{Declaration of generative AI and AI-assisted technologies in the writing process}

During the preparation of this work the authors used Grammarly and ChatGPT in order to improve language and readability. After using these tools/services, the authors reviewed and edited the content as needed and take full responsibility for the content of the publication.


\bibliographystyle{elsarticle-num}
\bibliography{references}

\end{document}